\documentclass[a4paper,twocolumn,11pt,accepted=2024-11-18]{quantumarticle}

\pdfoutput=1
\usepackage[T1]{fontenc}
\usepackage[utf8]{inputenc}
\usepackage{graphicx, amsmath, amssymb, dsfont, physics, xcolor, booktabs}
\usepackage[normalem]{ulem}
\usepackage[square,sort&compress,numbers]{natbib}

\graphicspath{{figures/}}

\definecolor{myblue}{HTML}{0000FF}
\definecolor{myred}{rgb}{1,0.,0.3}

\usepackage[bookmarks=true,colorlinks,linkcolor=quantumviolet,urlcolor=quantumviolet,citecolor=quantumviolet]{hyperref}

\begin{document}

\title{Flying Spin Qubits in Quantum Dot Arrays Driven by Spin-Orbit Interaction}

\author{D.~Fernández-Fernández}
\orcid{0000-0001-7163-8464}
\email{david.fernandez@csic.es}
\affiliation{Instituto de Ciencia de Materiales de Madrid (ICMM), CSIC, 28049 Madrid, Spain}
\author{Yue~Ban}
\orcid{0000-0003-1764-4470}
\affiliation{Departamento de Física, Universidad Carlos III de Madrid, Avda. de la Universidad 30, 28911 Leganés, Spain}
\author{G.~Platero}
\orcid{0000-0001-8610-0675}
\affiliation{Instituto de Ciencia de Materiales de Madrid (ICMM), CSIC, 28049 Madrid, Spain}

\begin{abstract}
	Quantum information transfer is fundamental for scalable quantum computing in any potential platform and architecture.
	Hole spin qubits, owing to their intrinsic spin-orbit interaction (SOI), promise fast quantum operations which are fundamental for the implementation of quantum gates.
	Yet, the influence of SOI in quantum transfer protocols remains an open question.
	Here, we investigate flying spin qubits mediated by SOI, using shortcuts to adiabaticity protocols, i.e., the long-range transfer of spin qubits and the quantum distribution of entangled pairs in semiconductor quantum dot arrays.
	We show that electric field manipulation allows dynamical control of the SOI, enabling simultaneously the implementation of quantum gates during the transfer, with the potential to significantly accelerate quantum algorithms.
	By harnessing the ability to perform quantum gates in parallel with the transfer, we implement dynamical decoupling schemes to focus and preserve the spin state, leading to higher transfer fidelity.
\end{abstract}

\maketitle

\section{Introduction}
In the late nineties, Loss and Divizenzo proposed semiconductor quantum dots (QDs) networks as a platform for a quantum computer \cite{Loss1998, Burkard1999}.
For almost a decade, the research focused on achieving one- and two-qubit operations in GaAs single and double QDs \cite{Ciorga2000, Petta2005, Nowack2007, Foletti2009} and on how to mitigate charge and spin decoherence in these systems \cite{Bluhm2010}, a critical concern in quantum computing.
Hereafter, the development of quantum dots arrays (QDAs) \cite{Vidan2004, Gaudreau2006, Schroeer2007, Rogge2008, Gaudreau2011, Noiri2022} marked a significant step toward scalability, opening the doors to new functionalities with applications in quantum computation, quantum information, and quantum simulation \cite{Creffield2002, Hensgens2017, PerezGonzalez2019a, Dehollain2020, Diepen2021, Xue2022}.

Triple quantum dots (TQDs) arranged linearly exhibit direct transport of electron spin states between outer dots, effectively using a QDA as a spin bus \cite{Busl2010, Busl2013, Sanchez2014, Sanchez2014a}.
In parallel, direct charge transfer between outer dots, which has been termed long-range transfer, has been observed in photoassisted tunneling experiments within a closed TQD system \cite{Braakman2013}.
Additionally, theoretical investigations have explored long-range photoassisted tunneling in TQDs \cite{GallegoMarcos2015, GallegoMarcos2016, Stano2015, PicoCortes2019, PicoCortes2021}.
\begin{figure}[!t]
	\centering
	\includegraphics[width=\linewidth]{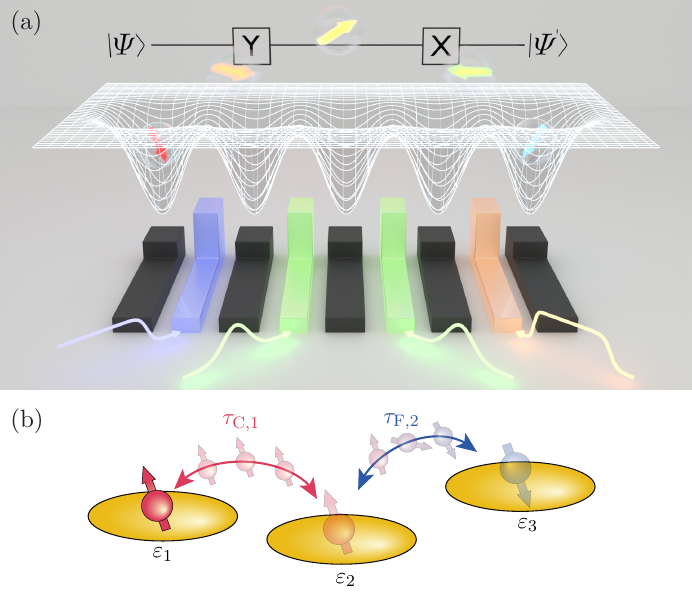}
	\caption{
		(a) Schematic picture of a linear five quantum dots array.
		Controlling the tunneling rates between the dots allows for the long-range transfer of spin qubits.
		Due to the presence of spin-orbit interaction, the transfer can be performed while simultaneously performing quantum operations.
		(b) Illustration of a triple quantum dot populated with a single particle.
		Each energy level of the dot is defined with the parameter $\varepsilon_i$.
		The particle can tunnel between adjacent dots $i$ and $i+1$ with a spin-conserving rate $\tau_{C, i}$.
		Due to the finite SOI, the particle can also tunnel to its neighboring dots with a spin-flip probability $\tau_{F, i}$.}
	\label{fig:schematic_3d}
\end{figure}
Recent years have witnessed the implementation of longer QDAs to increase the number of qubits and the complexity of quantum operations.
However, existing devices often lack high connectivity between qubits, a critical hurdle in simulating more complex systems.
To increase the number of problems that can be addressed with a quantum chip, the use of more dense architectures is desirable.
To address the challenge of signal fan-out \cite{Nguyen2017, Pauka2019}, reducing the number of input terminals through a crossbar network \cite{Li2018a, Boter2022} has been proposed as a solution.
Another approach to scalability is the use of sparse QDAs, eliminating qubit crosstalk and enabling integrated classical electronics \cite{Vandersypen2017}.
In this architecture, distant elements within a quantum chip are connected via quantum links or couplers.
Examples of them are electromagnetic cavities \cite{HarveyCollard2022}, surface acoustic waves \cite{Shilton1996}, bucket brigade \cite{Mills2019, RiggelenDoelman2024, Zwerver2023}, or conveyor modes \cite{Seidler2022, Struck2024, Langrock2023, Xue2024}, which facilitate the exchange of information between computing nodes \cite{Buonacorsi2019, Jnane2022, Kuenne2024}.

The framework proposed in this work allows the performance of quantum operations on qubits while they are coherently transferred, leading to "flying qubits" \cite{flyingQ-photon, flyingQ-electron2, flyingQ-electron1}, and the control of non-local entanglement with the advantages of scaling up over static qubits.
Already built based on quantum photonic \cite{flyingQ-photon} and electronic \cite{flyingQ-electron2, flyingQ-electron1} systems, a flying qubit architecture serves as a communication link for quantum computers \cite{flyingQ-QC} and secure data transmission for the quantum internet \cite{flyingQ-Qinternet}.
In this work, we investigate QDAs as quantum links between elements of a quantum chip, enabling coherent quantum information transfer via a flying spin qubit directly between the outer dots due to the presence of dark states (DSs) \cite{Greentree2004, Michaelis2006, Sanchez2013, Ban2018, Ban2019, Gullans2020}, as schematically shown in Fig.~\ref{fig:schematic_3d}~(a).
The use of long-range transfer protocols via DSs reduces drastically the population of the intermediate sites during the shuttling, decreasing its sensibility to noises occurring in this area.
Spin-orbit interaction (SOI) has been treated \cite{Bogan2017, Bogan2018, Studenikin2019, Bogan2019, Studenikin2021, Ducatel2021, Bogan2021, PadawerBlatt2022, Marton2023} as a source of decoherence during transfer \cite{Ginzel2020}, although it can be advantageous for achieving high-fidelity quantum gates \cite{Qi2023}.
We explore the long-range transfer of quantum information in QDAs under the influence of SOI which leads to a fast and robust entanglement distribution, a crucial element for quantum computing.
Furthermore, we propose a protocol for flying spin qubits, i.e., transferring spin qubits while simultaneously performing universal one-qubit gates.
Importantly, by controlling the effects of SOI, we demonstrate how to implement dynamical decoupling schemes alongside the transfer process, enhancing its fidelity.
We extend our analysis to investigate the effects of strong SOI on the long-range transfer where Coulomb interaction plays an important role.
We develop a rapid protocol for distributing entangled pairs in a QDA and perform long-range spin transfer in the half-filling regime.
Have already been implemented experimentally \cite{Liu2024}, Shortcuts to adiabaticity (STA) protocols are employed \cite{GueryOdelin2019} to efficiently accelerate adiabatic processes and enable high-fidelity transfer \cite{Ban2018, Ban2019}.

Our proposal simultaneously combines coherence, tunability, long-range transfer, and quantum gates implementation.
The existence of DSs is not limited to systems with a single particle, but also persist in systems where the interaction between particles is present.
Our results are valid for both electron \cite{Zajac2016, Otsuka2016, Baart2016, Fujita2017, Kandel2019, Qiao2020} spins and hole spin \cite{Lawrie2020, Yu2023} qubits in planar QDAs under high SOI in different semiconductor materials, such as germanium or silicon, and can also be extended to other semiconductor QDA implementations such as QDs in nanowires \cite{Fang2023}.
However, our primary focus is on hole spin qubits, where the intrinsic high SOI plays an important role \cite{FernandezFernandez2022}, allowing for rapid coherent spin rotations through electric dipole spin resonance \cite{FernandezFernandez2023}.

The manuscript is organized as follows.
In Sec.~\ref{sec:theoretical_model}, we introduce the theoretical model for the QDA and the SOI.
In Sec.~\ref{sec:dark_states}, we demonstrate how dark states arise in a TQD even in the presence of SOI and how they can be used for long-range transfer.
In Sec.~\ref{sec:transfer_protocol}, we discuss and compare different protocols for long-range spin qubit transfer.
In Sec.~\ref{sec:SOI_DS}, we illustrate how the addition of SOI modifies the DSs and how its control enables the implementation of quantum gates during the transfer process.
In Sec.~\ref{sec:applications}, we explore practical applications of the proposed quantum gates during the transfer, including the preparation of distant entangled pairs and the implementation of dynamical decoupling schemes.
In Sec.~\ref{sec:multiparticle_systems}, we explore two different scenarios with multiple particles in the QDA and demonstrate how DSs, in combination with SOI, play a crucial role in the long-range transfer of quantum information.
Finally, in Sec.~\ref{sec:conclusions}, we summarize our findings.

\section{Theoretical model}\label{sec:theoretical_model}
We consider a linear QDA (see Fig.~\ref{fig:schematic_3d}~(b)) with a total of $N$ sites, populated with heavy holes with nearest-neighbor couplings, described by an Anderson-Hubbard model ($\hbar = 1$) \cite{Burkard2023}
\begin{equation}
	H = H_0 + H_\tau + H_\mathrm{SOI}.
	\label{eq:total_hamiltonian}
\end{equation}

The original Hamiltonian reads
\begin{equation}
	H_0 = \sum_i \varepsilon_i n_i + U\sum_in_{i\uparrow}n_{i\downarrow} + \frac{E_Z}{2}\sum_i(n_{i\uparrow}-n_{i\downarrow}),
	\label{eq:original_hamiltonian}
\end{equation}
where $\varepsilon_i$ is the onsite energy of the dot $i$-th dot with $i=\left\{1, 2, \dots, N\right\}$, $U$ is the Coulomb repulsion on each dot, and $E_Z=g\mu_B B$ the Zeeman splitting given by external magnetic field.
The second term in Eq.~(\ref{eq:total_hamiltonian}) represents the spin-conserving tunneling between nearest-neighbor QDs as
\begin{equation}
	H_\tau = - \sum_{i, \sigma}\left(\tau_{C, i}a_{i\sigma}^\dagger a_{i+1\sigma} + \operatorname{h.c.}\right).
	\label{eq:tau_hamiltonian}
\end{equation}
Here, $a_{i\sigma}$ ($a_{i\sigma}^\dagger$) is the fermionic annihilation (creation) operator at site $i$ with spin $\sigma =\left\{\uparrow,\downarrow\right\}$, and $\tau_{C, i}$ the spin-conserving tunneling rate between the $i$-th and the $i+1$-th QDs.
For simplicity, we only consider one orbital per dot.
The last term in Eq.~(\ref{eq:total_hamiltonian}) models the SOI present for holes in semiconductor QDs.
We consider holes in Si or Ge planar QDs, which present cubic Rashba SOI \cite{Mutter2020, Bosco2021, Mutter2021b, Mutter2021a, Adelsberger2022, Jirovec2022}
\begin{equation}
	H_\text{SOI} = i\alpha(\sigma_+\pi_-^3-\sigma_-\pi_+^3).
	\label{eq:SOI_hamiltonian}
\end{equation}
The canonical momentum reads $\boldsymbol{\pi} = \mathbf{p}+e\mathbf{A}$, and the ladder operators are defined as $\pi_\pm = \pi_x \pm i\pi_y$ and $\sigma_\pm = (\sigma_x \pm i\sigma_y) / 2$, with $\sigma_{x,y,z}$ the Pauli matrices.
Following \cite{Mutter2021a}, we obtain the matrix elements for the spin-flip tunneling rates as
\begin{equation}
	\begin{split}
		\bra{i\uparrow}H_\text{SOI}\ket{i+1\downarrow} & = -\tau_{F,i},  \\
		\bra{i\downarrow}H_\text{SOI}\ket{i+1\uparrow} & = \tau_{F,i}^*.
	\end{split}
	\label{eq:SOI_terms}
\end{equation}
Other elements can be obtained by imposing hermiticity on the total Hamiltonian.
The SOI term can be written
in terms of the spin-flip tunneling rates between nearest neighbors as
\begin{equation}
	H_\mathrm{SOI} = \sum_{i}\left(\tau_{F, i}^*a^\dagger_{i\uparrow}a_{i + 1\downarrow}-\tau_{F, i}a^\dagger_{i\downarrow}a_{i + 1\uparrow} + \operatorname{h.c.}\right).
	\label{eq:phenom_SOI_hamiltonian}
\end{equation}

Without loss of generality, we restrict ourselves to real tunneling rates $\tau_{C, i}, \tau_{F, i} \in \mathds{R}$ (see Appendix~\ref{app:effective_spin_flip} for more details on the calculation of the spin-flip tunneling rate).
In the following sections, the shuttling protocol will be based on the dynamic control of the tunneling rates, both spin-conserving $\tau_{C, i}(t)$ and spin-flip $\tau_{F, i}(t)$.
It is important to note that the dynamical control of SOI via electrical field manipulation is experimentally feasible, as demonstrated in \cite{Nitta1997}.

\section{Dark states in a TQD with SOI}\label{sec:dark_states}
Our first aim is the long-range transfer of a single spin qubit prepared in a certain state across a linear QDA.
To reduce relaxation and dephasing effects, we investigate the direct transfer from the leftmost to the rightmost site, reducing as much as possible the population in the intermediate dots.
The minimal system for direct transfer is a TQD array populated with a single particle.

To obtain a long-range spin qubit transfer, we fix the QDs energy levels at $\varepsilon_i = 0$.
With no magnetic field applied to the system, the spin energy levels are degenerated.
We search for zero-energy modes that directly connect the edge dots of the chain with no population in the middle dot.
These coherent superpositions are termed dark states (DSs).
By exact diagonalization of the total Hamiltonian, and defining $x_\mathrm{SOI}$ as the ratio between spin-flip and the spin-conserving tunneling rates $x_\mathrm{SOI} \equiv\tau_{F, i} / \tau_{C, i}$ \cite{Mutter2021, Mutter2021b}, we find two DSs which read:
\begin{widetext}
	\begin{equation}
		\begin{split}
			\ket{\mathrm{DS}_1} = \sin\theta \ket{\uparrow, 0, 0} - \frac{\cos\theta}{x_\mathrm{SOI}^2+1}\left[(1-x_\mathrm{SOI}^2)\ket{0, 0, \uparrow} + 2x_\mathrm{SOI}\ket{0, 0, \downarrow}\right], \\
			\ket{\mathrm{DS}_2} = \sin\theta \ket{\downarrow, 0, 0} + \frac{\cos\theta}{x_\mathrm{SOI}^2+1}\left[(1-x_\mathrm{SOI}^2)\ket{0, 0, \downarrow} + 2x_\mathrm{SOI}\ket{0, 0, \uparrow}\right],
		\end{split}
		\label{eq:TQD_DSs}
	\end{equation}
\end{widetext}
where we have defined $\tan\theta \equiv \tau_{C, 2}/\tau_{C, 1}$.
At the beginning of the transfer protocol, the system is initialized in a DS by populating only the leftmost QD, i.e., $\tau_{C,1}\ll \tau_{C, 2}$.
The particle can be adiabatically transferred to the rightmost QD by tuning the ratio between the tunneling rates, until $\tau_{C,1} \gg \tau_{C,2}$.

It is important to note that for the correct creation of the DS, the leftmost QD and the rightmost QD must be in resonance.
Therefore, the only parameters that can drive long-range transfer using the DS are the tunneling rates $\tau_{C, i}$ and $\tau_{F, i}$, while the on-site energies $\varepsilon_i$ should remain fixed during the transfer process.

For simplicity, until otherwise stated, we will focus on $\ket{\mathrm{DS}_1}$ in Eq.~(\ref{eq:TQD_DSs}), but similar results can be obtained considering $\ket{\mathrm{DS}_2}$.
Note that any linear combination of $\ket{\mathrm{DS}_1}$ and $\ket{\mathrm{DS}_2}$ is also a zero-energy eigenvalue of the total Hamiltonian, so the initial state of the particle can be any superposition of spin up and spin down.

The initialization in an experimental setup should be conducted with large barriers between the dots, ensuring that the ground state of the system corresponds to a single isolated particle in the leftmost dot.
At this stage, the resonance condition between the dots can be relaxed.
After loading a single particle into the system, the QDs are brought into resonance, and the transfer protocol is initiated.

Due to systematic errors, small deviations from the ideal resonance condition can be expected in an experimental implementation.
In addition, the control of the SOI via a plunger gate can introduce modifications in the onsite energies, deviating from the ideal condition of resonance.
However, the DSs are robust against these errors, provided that first and last remains resonant.
In Appendix~\ref{app:dark_states_in_real}, we show that long-range transfer is still feasible with small deviations from the ideal resonance condition.

The requirement of zero magnetic field is crucial for the existence of DSs with a single particle in the presence of SOI, as it requires degeneracy between the spin states.
However, we will show in Sec.~\ref{sec:SOI_DS} that the addition of SOI can be used to control the final spin state of the particle after the transfer, enabling the implementation of a general one-qubit gate.
If the QDA is used as a quantum bus, connecting different computing nodes which require a non-zero magnetic field to operate, local magnetic fields through the entire quantum chip should be applied.
Other solution to circumvent this limitation is to introduce an AC electric field to drive photoassisted transitions between states with different spin projections, thereby recovering an effective resonance condition \cite{FernandezFernandez2023}.
When considering multiparticle systems, the presence of a high magnetic field is required for a high-fidelity long-range transfer, as shown in Sec.~\ref{sec:multiparticle_systems}.

\section{Comparison between transfer protocols}\label{sec:transfer_protocol}
In this section, we investigate different protocols to transfer a spin qubit in a TQD.
As shown in Eq.~(\ref{eq:TQD_DSs}), the existence of the DSs is independent of the present of a non-zero SOI.
To isolate the role of the DS in the long-range transfer, we fix the contribution from the SOI so that $x_\mathrm{SOI} = 1$, obtaining the dark state which reads
\begin{equation}
	\ket{\text{DS}_1(x_\mathrm{SOI}=1)} =\sin\theta \ket{\uparrow, 0, 0} - \cos\theta \ket{0, 0, \downarrow},
	\label{eq:DS1}
\end{equation}
where, by tuning $\theta$ from $\pi/2$ to $0$, the spin is inverted during the transfer.
The condition on $x_\mathrm{SOI}$ will be relaxed later to explore the effects of SOI on the transfer.

The system is initialized in the state $\ket{\Psi(t=0)}=\ket{\uparrow, 0, 0}$.
We compare each protocol through its fidelity, which is defined as the final population of the rightmost QD with a spin-down particle $\mathcal{F}\equiv|\braket{0, 0, \downarrow}{\Psi(T)}|^2$, where $T$ represents the total time of the protocol.
The maximum population of the middle dot is given by $P_2 \equiv \max(\sum_{\sigma} \abs{\braket{0, \sigma, 0}{\Psi(t)}}^2)$.

\begin{figure}[!t]
	\centering
	\includegraphics[width=\linewidth]{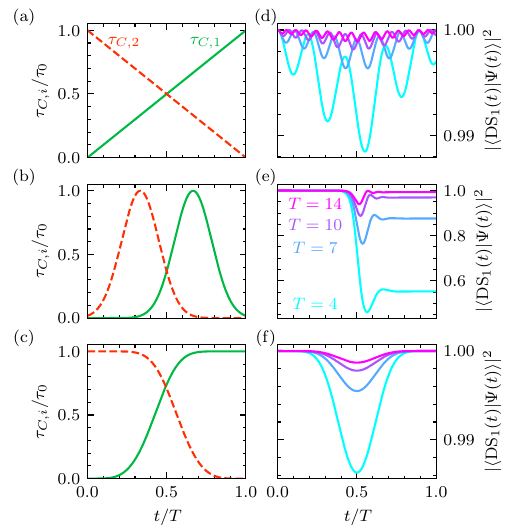}
	\caption{(a-c) Pulse shapes for (a) linear ramps, (b) CTAP, and (c) STA.
		(d-f) Overlap between the dark state $\ket{\mathrm{DS}_1(t)}$ and the time-dependent wave function $\ket{\Psi(t)}$ during the state transfer implemented with (d) linear ramps, (e) CTAP, and (f) STA, applied in a TQD.
		The colors for the right column represent the total time of the protocol in units of $[2\pi/\tau_0]$, shown in (e).
		Note that the $y$ axis scale for the right column is different for each protocol.}
	\label{fig:1HH_transfer_overlapping_DS}
\end{figure}

One of the simplest protocols for the transfer of a single particle consists of two linear ramps at tunneling rates given by
\begin{equation}
	\begin{split}
		\tau_{C, 1}(t) & = \frac{\tau_0}{T}t,          \\
		\tau_{C, 2}(t) & = \tau_0 - \frac{\tau_0}{T}t, \\
	\end{split}
\end{equation}
where $\tau_0$ represents the maximum tunneling rate.
An illustrative example of pulse shapes can be seen in Fig.~\ref{fig:1HH_transfer_overlapping_DS}~(a).
It is worth noting that spin-flip tunneling rates also evolve with time, following $\tau_{F, i}(t) = x_\mathrm{SOI}\tau_{C, i}(t)$.

Another protocol of great interest in the literature for quantum control is coherent transfer by adiabatic passage (CTAP) \cite{Greentree2004, Ban2018}.
In CTAP, pulses are defined by Gaussian shapes as follows
\begin{equation}
	\tau_{C,1(2)} = \tau_0 \exp(-\frac{(t-T/2 \mp \sigma_\mathrm{CTAP})^2}{\sigma_\mathrm{CTAP}^2}),
\end{equation}
where $\sigma_\mathrm{CTAP}$ is a free parameter that determines the standard deviation of the control pulse, with the pulse shape depicted in Fig.~\ref{fig:1HH_transfer_overlapping_DS}~(b).

To enhance long-range transfer, we explore STA schemes \cite{Chen2010, GueryOdelin2019}, which speeds up adiabatic protocols increasing the transfer fidelity.
In this work, we adopt the inverse engineering framework.
We propose an ansatz for the state evolution and analytically solve the time-dependent Schrödinger equation to obtain the pulse shapes that should be applied for the transfer.
Crucially, our ansatz allows for the controlled population in the middle dot at intermediate times, with the maximum population tuned through the pulse strength.
The ansatz for the wave function reads
\begin{equation}
	\begin{split}
		\ket{\Psi(t)} = & \cos\eta\cos\chi \ket{\uparrow, 0, 0} -\sin\chi\cos\eta\ket{0, 0, \downarrow} \\
		                & + i\frac{\sin\eta}{\sqrt{2}}(\ket{0, \uparrow, 0}+\ket{0, \downarrow, 0}),
	\end{split}
	\label{eq:ansatz}
\end{equation}
where $\eta$ and $\chi$ are some auxiliary time-dependent functions.
We will see later that the pulses obtained with IE are valid for arbitrary values of $x_\mathrm{SOI}$.
By introducing Eq.~(\ref{eq:ansatz}) into the time-dependent Schrödinger equation $i\partial_t\ket{\Psi(t)}=H(t)\ket{\Psi(t)}$ we obtain the following expressions
\begin{equation}
	\begin{split}
		\tau_{C, 1}(t) = (\dot{\eta}\cos\chi + \dot\chi\cot\eta\sin\chi) / \sqrt{2}, \\
		\tau_{C, 2}(t) = (-\dot{\eta}\sin\chi + \dot{\chi}\cot\eta\cos\chi) / \sqrt{2}.
		\label{eq:pulses_1HH}
	\end{split}
\end{equation}
The boundary conditions for the required initial and final states are $\eta(0) = \eta(T) = \chi(0) = 0$, and $\chi(T)=\pi/2$.
We can also impose smooth pulses by using additional boundary conditions in derivatives $\dot{\chi}(0) = \dot{\chi}(T) = \ddot{\chi}(0) = \ddot{\chi}(T) = \dot\eta ( 0) = \dot\eta (T) = 0$.
We use a common choice for smooth pulses, a Gutman 1-3 trajectory, which consists of a linear term and the two lowest odd Fourier components.
The explicit expressions for the auxiliary functions are
\begin{equation}
	\begin{split}
		\chi(t) = & \pi\frac{t}{2 T} - \frac{1}{3}\sin(\frac{2\pi t}{T}) + \frac{1}{24}\sin(\frac{4\pi t}{T}), \\
		\eta(t) = & \arctan(\dot\chi / \alpha_0),
		\label{eq:auxiliary_functions}
	\end{split}
\end{equation}
where $\alpha_0$ is an arbitrary constant controlling both the maximum pulse strength $\tau_{0} \propto \alpha_0$ and the total population in the middle dot $P_2(t) \equiv \sin^2\eta(t)$.
An example of pulse shapes is shown in Fig.~\ref{fig:1HH_transfer_overlapping_DS}~(c).

The coefficients for each Fourier component in Eq.~(\ref{eq:auxiliary_functions}) can be fine-tuned to minimize the population in the intermediate dot.
Employing more Fourier components enhances transfer fidelity at the cost of more intricate pulse shapes.
However, optimizing these parameters requires an exhaustive search and falls outside the scope of this work.

The maximum population in the middle dot occurs at the midpoint of the transfer and is given by
\begin{equation}
	P_2\equiv P_2(T/2) = \frac{(4\pi/3)^2}{(4\pi/3)^2 + T^2\alpha_0^2}.
\end{equation}
When the total protocol time is significantly short, compared to the maximum pulse strength $T\alpha_0\ll 1$, the middle dot is maximally populated $P_2\rightarrow 1$.
In such a case, the adiabatic condition cannot be fulfilled, but the precise engineering of pulses obtained with IE ensures that the system ends in the desired final state.
To reduce the middle dot population, longer total times or stronger pulses must be used.

We compute the evolution with the total Hamiltonian given in Eq.~(\ref{eq:total_hamiltonian}), accounting for the time-dependent tunneling rates dictated by each protocol.
Subsequently, we calculate the overlap between the time-dependent wave function and $\ket{\mathrm{DS}_1(t)}$ during the transfer and define the fidelity as the overlap at $t=T$.
The results for different total times are depicted in Fig.~\ref{fig:1HH_transfer_overlapping_DS}~(d-f).
Since all protocols start with $\tau_{C, 2}\gg \tau_{C, 1}$, the initial state $\ket{\uparrow, 0, 0}$ corresponds to the DS in the limit $t\rightarrow0$, which produces an overlap close to one at early times.

Fig.~\ref{fig:1HH_transfer_overlapping_DS}~(d) shows that by using linear ramp pulses, the overlap remains close to one for all total times examined here.
The most significant deviation from the DS during the transfer occurs in the middle of the protocol, around $t \sim T/2$.
This behavior is also observed in all the other protocols.
As the dynamics approaches the adiabatic limit at longer times, the state remains in the DS, reducing the total middle dot population and achieving complete transference to the third QD.
However, even if the overlap is close to one, the oscillatory behavior of the overlap makes linear ramps highly sensitive to the total transfer time.
A slight deviation in time results in substantial differences in transfer fidelity.
When using CTAP pulses (Fig.~\ref{fig:1HH_transfer_overlapping_DS}~(e)), the overlap with the DS drops in the middle of the protocol, but it recovers at later times.
If the imposed transfer time is too short, the drop is too severe, making it impossible to achieve high-fidelity transfers.
On the other hand, the STA protocol (Fig.~\ref{fig:1HH_transfer_overlapping_DS}~(f)) behaves qualitatively similar to CTAP for shorter times $t<T/2$.
However, due to the precise engineering of the pulses, STA consistently achieves a perfect overlap with the DS at the end of the protocol, regardless of the total time of the protocol.
This results in high robustness against timing errors.

DS-mediated state transfer extended to arrays of more than three sites provided that the QDA has an odd number of sites \cite{Greentree2004, Ban2018, Ban2019}.
The transfer for $N>3$ can be obtained using the tunneling sequence of straddling pulses \cite{Malinovsky1997, Vitanov2001a, Greentree2004}.
This sequence consists of a pulse in the tunneling rates between the first and second sites and between the last two sites, as for the TQD case.
The barriers between the middle dots, termed bulk barriers, are raised and lowered simultaneously with a time-dependent function such as $\tau_{C, i}=\tau_s\exp(-(t-T/2)^2/(2\sigma_\mathrm{bulk}^2))$ with $1<i<N-1$ and $\tau_s$ being the maximum value for straddling pulses.
If we impose large straddling pulses $\tau_s\gg\tau_0$, the system can be reduced to a three-site system with renormalized tunneling rates $\tau_{C, 1}' = -\tau_{C, 1}/\sqrt{(N-1)/2}$ and $\tau_{C, 2}' = (-1)^{(N-1)/2}\tau_{C, N-1}/\sqrt{(N-1)/2}$, and similarly, with renormalized spin-flip tunneling rates $\tau_{F, 1}'$ and $\tau_{F,2}'$ (see Appendix~\ref{app:larger_arrays} for more information about how to obtain the effective model).
These two pulses are usually referred to as pump and Stokes pulses, respectively.

To benchmark the long-range transfer protocols discussed above, we compare them with the traditional sequential transfer protocol, also known as the bucket-brigade mode, where the tunneling rates between all neighboring dots are maintained at a constant value, and the driving parameter is the detuning between QDs.
Initially, the particle is trapped in one QD due to a significant detuning from neighboring dots.
Subsequently, the detuning is gradually reduced until the ground state corresponds to the particle being trapped in the neighboring dot.
If this ramp is sufficiently slow (adiabatic), the particle is sequentially transferred from one QD to its neighbor, reaching the final site after a total of $N - 1$ sequences.
Here, we define the maximum value of the detuning as $\varepsilon_\mathrm{max}$, and the minimum value at $\varepsilon_\mathrm{min}=0$.

\begin{figure}[tb!]
	\centering
	\includegraphics[width=\linewidth]{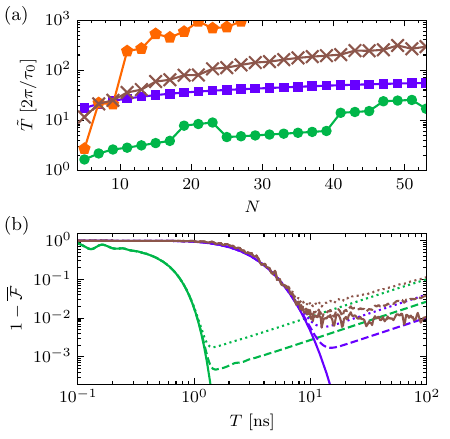}
	\caption{
		(a) Time needed to reach a $99\%$ fidelity for different protocols, versus the number of sites in the QDA.
		The transfer is obtained by sequential transfer (crosses, brown), linear pulse (pentagons, orange), CTAP (squares, purple), and STA (circles, green).
		The maximum straddling tunneling rate is $\tau_s = 3\tau_0$.
        For sequential transfer $\varepsilon_\mathrm{max} = 20\tau_0$.
		(b) Transfer infidelity as a function of the total protocol time for different protocols for a seven QDA.
		The value of the dephasing is $\gamma=0$ (solid), $\gamma=0.05\; \mu$eV (dashed), and $\gamma=0.1\; \mu$eV (dotted).
		Maximum tunneling rates are $\tau_s = 10 \tau_0 = 100\; \mu$eV.
		For sequential transfer $\varepsilon_\mathrm{max} = 200\; \mu$eV.
		Other parameters, common for both panels, $\sigma_\mathrm{CTAP}=\sigma_\mathrm{bulk}=T/6$.
	}
	\label{fig:protocols_comparison}
\end{figure}

We define $\tilde{T}$ as the minimal protocol time needed to obtain a quantum state transfer with fidelity of $\mathcal{F}=99\%$.
In Fig.~\ref{fig:protocols_comparison}~(a), we show the results for different protocols versus the number of sites in the QDA.
The total time needed using a linear ramp grows exponentially with the number of sites.
On the contrary, an STA pulse reaches a high-fidelity transfer with moderate times.
For very large arrays, the CTAP and STA protocols approach each other.
Since the straddling pulses in the bulk are the same for both protocols, this result is expected.
However, this assumption can not be extrapolated to linear pulses, where the time necessary to achieve a $99\%$ fidelity abruptly increases with the length of the chain as compared with the other protocols.
Both STA and CTAP outperform sequential transfer, demonstrating the superiority of employing long-range transfer in long chains.
Furthermore, as shown in Fig.~\ref{fig:protocols_comparison}~(a), STA is much faster than CTAP for intermediate QDA lengths, making it the most efficient protocol for long-range transfer.

Another advantage of long-range transfer is the reduction of the effects of pure dephasing compared to sequential transfer.
To simulate this effect, we solve the Lindblad master equation \cite{Manzano2020} with jump operators $L_i = \sqrt{\gamma_i}\sigma_z^i$, where $\gamma_i$ is the dephasing rate for the $i$-th QD.
We assume the same dephasing rate $\gamma_i = \gamma$ for all quantum dots and compute the fidelity $\overline{\mathcal{F}}$ as the average of the transfer fidelity for different initial spin polarization.
Fig.~\ref{fig:protocols_comparison}~(b) shows the fidelity as a function of the dephasing rate $\gamma$ for different protocols.
Since long-range transfer protocols do not populate the intermediate dots, the fidelity is less affected by dephasing, as compared with other protocols, obtaining a higher fidelity for the same dephasing rate.
The advantage of STA or CTAP over sequential transfer is even more evident for higher dephasing rates and longer QDAs (not shown).
In Appendix~\ref{app:charge_noise} we give a brief discussion on the effects of charge noise on the transfer fidelity for different protocols.

Finally, it is important to note that for the transfer, it is crucial to maintain the resonance condition between the first and last QDs.
However, due to the long-range nature of the DS, the transfer fidelity is not sensitive to the presence of small local magnetic fields in the intermediate dots.
We have verified numerically that a magnetic field with a Zeeman splitting $E_Z \lesssim \tau_0$ in the intermediate dots does not drastically affect the transfer fidelity.
For instance, for an $N=11$ QDA with other parameters given by the ones used in Fig.~\ref{fig:protocols_comparison}{(a)}, the fidelity drop from $E_Z=0$ to $E_Z=\tau_0$ is less than $1\%$ for the STA protocol.

\section{Role of SOI in the dark state}\label{sec:SOI_DS}
\begin{figure}[!t]
	\centering
	\includegraphics[width=\linewidth]{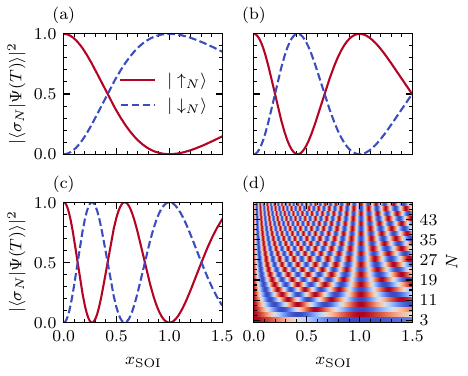}
	\caption{
		The population of the rightmost dot after a long-range transfer as a function of $x_\mathrm{SOI}=\tau_{F, i} / \tau_{C, i}$ for (a) $N=3$, (b) $N=5$, and (c) $N=7$.
		The spin is initialized at the left-most QD with spin up, and it is transferred to the right-most QD ending with spin up (solid red line), or with spin down (dashed blue line).
		(d) Spin polarization at the right-most QD at the end of the protocol as a function of $x_\mathrm{SOI}$ and $N$ is an odd number (blue: spin down, red: spin up).
		The protocol used for the pump and Stokes pulses is an STA pulse, with a maximum value of $\max(\tau_{C, 1})=\max(\tau_{C, N - 1})=\tau_0\sqrt{(N - 1) /2}$, while the intermediate tunneling rates are straddling pulses with Gaussian shapes centered at $T/2$, with width $\sigma_\mathrm{bulk}=T/6$ and maximum strength of $\tau_s=15\tau_0\sqrt{(N - 1) /2}$.
		The total protocol time is $T=20\pi/\tau_0$.}
	\label{fig:x_run_N}
\end{figure}

Let us perform a detailed analysis of DSs in a chain with an odd number of QDs, $N = 2k + 1$, where $k\in \mathds{N}$.
In the limit $\tau_s \gg \tau_0$, we derive a solution as follows:
\begin{equation}
	\begin{split}
		\ket{\mathrm{DS}_1} & = \sin\theta \ket{\uparrow_1} - \cos\theta\left[\cos\vartheta/2\ket{\uparrow_N}\right.\\
		&\left.\hspace{3.3cm}+\sin\vartheta/2\ket{\downarrow_N}\right],    \\
		\ket{\mathrm{DS}_2} & = \sin\theta \ket{\downarrow_1} - \cos\theta\left[-\sin\vartheta/2\ket{\uparrow_N}\right.\\
		&\left.\hspace{3.3cm}+\cos\vartheta/2\ket{\downarrow_N}\right].
	\end{split}
	\label{eq:general_DS}
\end{equation}
Here, the mixing angle between spin states at the last site is given by
\begin{equation}
	\vartheta \equiv 2(N - 1)\arctan(x_\mathrm{SOI}).
	\label{eq:rotation_angle}
\end{equation}
This outcome reveals the dependence of the spin polarization of the transferred particle on the number of QDs in the array and on the strength of the SOI, which can be modulated through external electric fields.
For the case of $N=3$, the spin is inverted in the transfer for $x_\mathrm{SOI} = 1$.
On the contrary, spin-conserving transfer occurs at $x_\mathrm{SOI}=0$, where the SOI is zero.
In Fig.~\ref{fig:x_run_N}~(a), simulations of the spin polarization at the rightmost QD after a long-range transfer carried out with STA are shown.
The values of $x_\mathrm{SOI}$ that yield long-range spin-conserving and spin-flip transfers are in agreement with the analytical results discussed above.

For $N=5$, solutions for spin-flip transfer ($\vartheta = (2k + 1)\pi$) emerge at $x_\mathrm{SOI}=\sqrt{2}\pm 1$.
On the other hand, for a spin-conserving transfer ($\vartheta = 2k\pi$), there are three possible solutions.
These correspond to the absence of SOI $x_\mathrm{SOI}=0$, equal spin-conserving and spin-flip tunneling rates $x_\mathrm{SOI}=1$, and absence of spin-conserving tunneling rate $x_\mathrm{SOI}\rightarrow \infty$.
Similar results are expected for higher values of $N$.
These predictions are closely aligned with the numerical results presented in Fig.~\ref{fig:x_run_N}~(b-c).

In addition, we simulate various chain lengths for long-range transfer using an STA protocol and compute the final polarization defined as $\mathcal{P}\equiv |\braket{\uparrow_N}{\Psi(T)}|^2-|\braket{\downarrow_N}{\Psi(T)}|^2$.
As in previous cases, the system is initialized in the state $\ket{\uparrow_1}$.
The results are displayed in Fig.~\ref{fig:x_run_N}~(d), where blue regions ($\mathcal{P}=-1$) represent spin-flip transfers, while red ones ($\mathcal{P}=1$) indicate spin-conserving ones.
The analysis is limited to chains with an odd number of QDs since only these systems, in the case of a single particle, exhibit dark states.
As described in Appendix~\ref{app:larger_arrays}, the effective tunneling rates decrease as $1 / \sqrt{(N -1) / 2}$ for increasing site number.
To counteract this effect, we increase the pulse strength by a factor $\sqrt{(N -1) /2}$, ensuring that the effective tunneling rate remains consistent across all QDA.
At $x_\mathrm{SOI}=1$, we observe the alternating behavior of the spin polarization with the number of QDs mentioned above.
The results resemble an interference pattern with an increasing fringe frequency as the chain length increases.

\subsection{Simultaneous one-qubit gate and quantum transfer}\label{sec:one_qubit_gate}
As we have discussed previously, the presence of SOI produces the spin rotation of the particle during the transfer.
One can tune this rotation by controlling the effective Rashba interaction.
In the present setup, the spin rotation takes place around the $y$ axis in the Bloch sphere.
Consequently, the final unitary transformation, after tracing out the spatial part, can be represented as
\begin{equation}
	R_Y(\vartheta) = \mqty(\cos\vartheta/2 & -\sin\vartheta/2 \\ \sin\vartheta/2 & \cos\vartheta/2).
\end{equation}

\begin{figure}[!t]
	\centering
	\includegraphics[width=\linewidth]{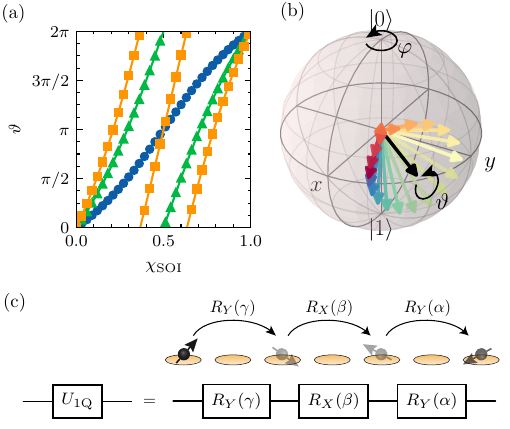}
	\caption{
		(a) Spin rotation angle versus the parameter $\chi_\mathrm{SOI}$.
		Long-range transfer is obtained using STA pulses in a QDA with a total of $N=3$ (blue, circles), $N=5$ (green, triangles), and $N=7$ (orange, squares) sites.
		Both numerical (markers) and analytical results (solid lines) are shown.
		(b) Sketch of the rotation angles in the Bloch sphere.
		The rotation angle $\vartheta$ is given by the SOI parameter $\chi_\mathrm{SOI}$, while the rotation vector is defined via the azimuth angle $\varphi$.
		This angle is modified by the driving frequency of the Rashba term.
		(c) An universal one-qubit gate can be implemented by combining a maximum of three rotations around two perpendicular axes.
		This can be obtained by dividing the total QDA into three subarrays, so each rotation is performed in a different subarray.
	}
	\label{fig:rotation_modified}
\end{figure}

To determine the final rotation angle, the time-dependent Schrödinger equation is solved numerically.
In Fig.~\ref{fig:rotation_modified} (a), we compare these numerical results with the analytical prediction provided in Eq.~(\ref{eq:rotation_angle}), showing an excellent agreement.
Here, we introduce the parameter $\chi_\mathrm{SOI}$, defined as $\chi_\mathrm{SOI} \equiv (1/x_\mathrm{SOI} + 1)^{-1}$.

To enable arbitrary one-qubit gates, it is imperative to have the capability for two-axis control.
This requirement can be achieved by introducing an effective SOI with a complex phase.
In this case, the DSs take the form
\begin{equation}
	\begin{split}
		\ket{\mathrm{DS}_1} & = \sin\theta \ket{\uparrow_1} - \cos\theta\left[\cos \vartheta/2\ket{\uparrow_N}\right.                                               \\
		                    & \phantom{\sin\theta \ket{\uparrow_1} - \cos\theta\left[\right]}\left.+e^{i\varphi}\sin \vartheta/2\ket{\downarrow_N}\right],  \\
		\ket{\mathrm{DS}_2} & = \sin\theta \ket{\downarrow_1} - \cos\theta\left[-\sin\vartheta/2\ket{\uparrow_N}\right.                                             \\
		                    & \phantom{\sin\theta \ket{\downarrow_1} - \cos\theta\left[\right]}\left.+e^{i\varphi}\cos\vartheta/2\ket{\downarrow_N}\right].
	\end{split}
\end{equation}

As discussed in \cite{Nitta1997, Malshukov2003, Faniel2011, Li2018}, to apply a periodic electric field to the sample, implies that the Rashba SOI acquires a complex phase as $\alpha=\alpha_0e^{i(\omega t + \phi)}$.
This also implies a time dependent spin-flip tunneling rate $\tau_\mathrm{F,i}(t)=\abs{\tau_{F, i}(t)}e^{i(\omega t + \phi)}$, where $\omega$ is the frequency and $\phi$ the phase of the drive.
Depending on the specific pulse protocol considered (STA, CTAP, or linear), the SO amplitude $\alpha_0$ becomes time-dependent in a particular fashion.

The effect of periodic driving on the spin-flip tunneling rate allows the control of the rotation axis through the frequency and phase of the driving.
If the driving frequency is sufficiently low, the particle dynamics follows the DS, and the final phase between the spin-up and spin-down states is determined by $\varphi \equiv \omega T + \phi$.
For $\varphi=0$, the rotation occurs around the $x$-axis, while for $\varphi=\pi/2$ the rotation takes place around the $y$-axis.
Here, we only mention these two possibilities, but any rotation vector in the $x$-$y$ plane can be chosen by modifying the driving phase.
A schematic representation is provided in Fig.~\ref{fig:rotation_modified}~(b).

The control of the SOI can be obtained through a global electric field applied to the sample, while plunger gates on top of the QDs can be used to locally control the tunneling rates, compensating for any deviations produced by the control of the SOI.
Note that the gate needed to control the SOI is not depicted in the schematic representation of the QDA in Fig.~\ref{fig:schematic_3d}~(a).

Combining three rotations around two perpendicular axes, any one-qubit gate can be implemented \cite{Barenco1995}.
Moreover, if the rotation axes can be adjusted freely within a set plane, such as the $x$-$y$ plane, then achieving a one-qubit gate only requires two rotation steps \cite{Shim2013}.
In Fig.~\ref{fig:rotation_modified}~(c), we show a possible implementation of a universal one-qubit gate.
The total QDA is divided into three subarrays, each containing three QDs.
The first and third subarrays are used to implement rotations around the $y$-axis, while the second subarray is used to implement a rotation around the $x$-axis.
Usual one-qubit gates times for spin qubits $T_\mathrm{1Q}\sim 10-100$ ns \cite{Stano2022}, both for electrons and holes, are comparable with the timescales of the long-range transfer protocols presented here.

\section{Applications for simultaneous quantum gates during transfer}\label{sec:applications}
In the following section, we will discuss two applications of quantum gates implemented simultaneously than the long-range transfer.
These applications motivate the use of the SOI to speed up the generation of entangled pairs and to enhance the robustness of the transfer, respectively.

\subsection{Quantum entanglement distribution}
\begin{figure}[!t]
	\centering
	\includegraphics[width=\linewidth]{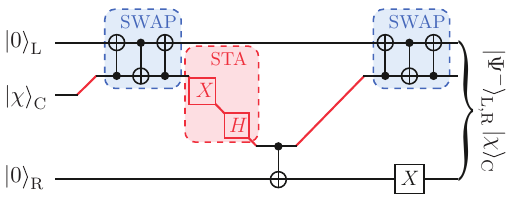}
	\caption{Quantum circuit for a distributed quantum state preparation mediated by a quantum bus.
	The outer qubits are initialized at $\ket{0}_\mathrm{L, R}$, while the initial state of the qubit inside the quantum bus (ancilla qubit) is arbitrary $\ket{\chi}_\mathrm{C}$, and it is transferred via a long-range protocol (red lines).
	During the second transfer, two one-qubit gates ($X$ and $H$) are applied to the ancilla qubit, speeding up the protocol.
	The final state for the outer qubits is the Bell pair $\ket{\Psi^-}_\mathrm{L, R} \equiv (\ket{0}_\mathrm{L}\otimes\ket{1}_\mathrm{R}-\ket{1}_\mathrm{L}\otimes\ket{0}_\mathrm{R})/\sqrt{2}$, and the ancilla qubit recovers its initial state.
	}
	\label{fig:aplication_1Q_gate}
\end{figure}

The use of simultaneous one-qubit gates in parallel with the transfer of quantum information is also advantageous for the distribution of quantum entanglement, see Fig.~\ref{fig:aplication_1Q_gate}.
The goal is to generate entanglement between two distant qubits that are not directly coupled to each other.
However, they are coupled to a third qubit, which can be shuttled between them, using the long-range transfer protocols discussed above.
Due to the non-zero SOI, one can use the second shuttling depicted in Fig.~\ref{fig:aplication_1Q_gate} to perform two quantum gates simultaneously to the transfer, speeding up the protocol.
The final Bell pair between the distant qubits can be chosen via the quantum gates performed during the shuttling, while the ancilla qubit recovers its original quantum state after the protocol.
In this example, outer qubits represent distant quantum cores, which are operated individually, and the ancilla qubit would denote a particle inside a quantum bus connecting both cores.

\subsection{Dynamical decoupling}\label{sec:dynamical_decoupling}
In experimental devices there exists a small but finite hyperfine interaction (HFI) that can induce spin decoherence and relaxation.
We model the HFI as a quasistatic Gaussian noise, which gives rise to local fluctuating magnetic fields characterized by a zero mean and a standard deviation of $\sigma_\mathrm{hf}$.
The timescale of the hyperfine interaction is long enough so that the random magnetic fields remain constant during the transfer.
These fluctuating magnetic fields introduce errors in the transfer process and also induce dephasing in the moving spin qubit.

To improve the robustness of the protocol, we can leverage the opportunity to apply quantum gates in parallel with the transfer.
We show below that it is possible to implement a dynamical decoupling (DD) protocol while transferring, and, as a proof of concept, we consider the Hahn echo protocol.
The Hahn echo protocol involves a sequence of three pulses around one rotation vector.
It has already been implemented in standing spin qubits in semiconductor QDs \cite{Huang2019, Hendrickx2021, RiggelenDoelman2024}.
In our setup, the natural rotation axis is the $y$-axis.
The first and third quantum gates introduce a rotation angle of $\pi/2$, while the second gate is characterized by a rotation angle of $\pi$.
Ideally, after the complete sequence, the qubit returns to its initial state.
A schematic representation of the dynamical decoupling protocols explained below is shown in Appendix~\ref{app:dynamical_decoupling}.

In our system, we divide the total QDA into a total of three subarrays of length $n_3$ each, to perform long-range transfers.
The total length of the QDA is given by $N = 3\times n_3 - 2$, accounting for the two QDs shared between adjacent subarrays.
During each transfer, the rotation angle is fixed by tuning the SOI.
We refer to this method as a three-steps DD.

However, when very long arrays are considered, the increasing number of dots in the subarrays can lead to lower quantum gate fidelity, and to address this issue, we propose an alternative approach.
The total QDA is divided into five subarrays.
The first, third, and fifth subarrays consist of three sites each, in which we apply the quantum gates required by the Hahn echo protocol.
The remaining subarrays, the second and fourth, consist of any odd number of sites $n_5$, in which the identity quantum gate are applied.
The value of $x_\mathrm{SOI}$ in this region is chosen so that the spin is not rotated during the transfer ($\vartheta = k\pi$).
Any possible errors during the shuttlings in the second and fourth subarrays are corrected by the dynamical decoupling protocol applied in the smaller subarrays.
The total length of the array is given by $N = 3\times 3 + 2 \times n_5 - 4$, so this approach is termed five-steps DD.
Ideally, the intermediate subarrays are chosen to be as long as possible, with $n_5\gg 3$.

To benchmark these protocols, we compare their results with those obtained by dividing the total QDA similarly but without applying quantum gates to the transferred qubit.
The minimum length of the QDA that can be used to compare both proposals is $N=19$.
For all long-range transfers during the dynamical decoupling schemes, we use the STA protocol, since it is the fastest one.

\begin{figure}[!t]
	\centering
	\includegraphics[width=\linewidth]{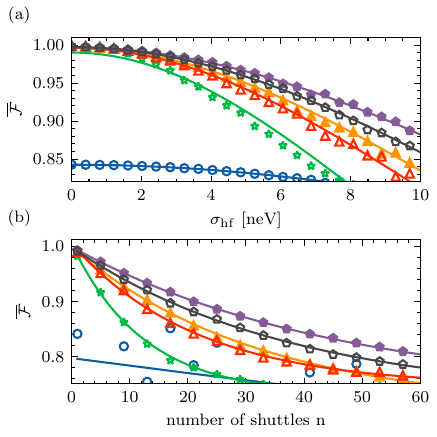}
	\caption{
		Average transfer fidelity for sequential (blue, circles), direct (green, stars), 3-steps without DD (red, open triangles), 3-steps with DD (orange, full triangles), 5-steps without DD (gray, open pentagons), and 5-steps with DD (violet, full pentagons).
		Each point represents an average over multiple random initial spin states and a random initialization of the Gaussian noise.
		In panel (a), transfer fidelity as a function of spin dephasing for a single shuttling.
		In panel (b), multiple forward and reverse shuttlings are performed, while the noise strength is fixed at $\sigma_\mathrm{hf} = 2$~neV.
		Solid lines represent fits for the numerical data, see main text.
		For both panels, $N = 19$, $\tau_0 = 1\; \mu$eV, and $T \sim 300$~ns.}
	\label{fig:dynamical_decoupling}
\end{figure}

The results presented in Fig.~\ref{fig:dynamical_decoupling} (a-b) demonstrate the effectiveness of various transfer protocols.
Here, we define direct transfer as the long-range transfer between outer dots with a SOI so that the transferred qubit remains in the same polarization as the initial state.
The spin dephasing time is given by $T_2^*=\hbar/\sigma_\mathrm{hf}$ \cite{Assali2011}, and the values considered here are in line with the dephasing time observed in hole spin qubits \cite{Stano2022}.
We conducted multiple shuttle operations, initializing the particle in the left QD with a random spin state and averaging the transfer fidelity over different realizations of random hyperfine interactions.
In particular, the fidelity obtained with sequential passages (see blue line in Fig.~\ref{fig:dynamical_decoupling}~(a)) remains low even in the absence of hyperfine interaction.
However, this protocol exhibits increased robustness against errors because it does not rely on the energy levels being in resonance, as in the case where the protocol is based on the existence of DSs.

The fidelity for the direct transfer protocol, while achieving high values in the limit of low hyperfine strength, decreases substantially as the hyperfine strength increases.
As we show in Fig.~\ref{fig:dynamical_decoupling} (open triangles and open pentagons), dividing the total QDA into smaller subarrays enhances robustness.
Furthermore, applying the dynamical decoupling sequence simultaneously to the transfer yields significant improvements in fidelity (full triangles and full pentagons).
These results highlight the advantage of executing quantum gates simultaneously to quantum state transfer.

To quantitatively assess the performance of dynamical decoupling protocols, we fitted the fidelity to an exponential decay of the form \cite{Struck2024}
\begin{equation}
	\overline{\mathcal{F}} = \frac{\mathcal{F}_0}{2}\left[1 + \exp(-(\sigma_\mathrm{hf} / \sigma_\mathrm{hf}^*)^2)\right],
	\label{eq:dd_hf}
\end{equation}
where $\mathcal{F}_0$ and $\sigma_\mathrm{hf}^*$ are the fitting parameters.
The value of $\sigma_\mathrm{hf}^*$ is indicative of the robustness of the protocol.
In a real-case scenario, the particle must experience multiple shuttlings to connect distant quantum computing nodes, as shown in the previous example.
To study the performance of the DD protocols in this scenario, we consider multiple forward and reverse shuttlings.
Here, we fit the numerical data to an exponential decay of the form \cite{RiggelenDoelman2024}
\begin{equation}
	\overline{\mathcal{F}} = \mathcal{F}_0\exp(-n / n^*) + \mathcal{F}_\mathrm{sat},
	\label{eq:dd_n}
\end{equation}
where $n^*$ defines the number of shuttles after which fidelity drops to $1/e$ of its initial value, and $\mathcal{F}_\mathrm{sat}$ is the saturation value of the fidelity.

The results for the fitting parameters are shown in Table~\ref{tab:dd_parameters}, where we can see that the application of the DD sequence significantly improves the robustness.
Our results show that by including a DD scheme during the shuttling, the qubit transfer is more robust against hyperfine interaction, and a larger number of shuttles can be performed before the fidelity significantly drops.

\begin{table*}[thb!]
	\begin{tabular}{@{}c|c|c|c|c|c|c@{}}
		\toprule
		                             & Sequential & Direct & Three-steps & Three-steps (DD) & Five-steps & Five-steps (DD) \\ \midrule
		$\mathcal{F}_0$              & 0.842      & 0.990  & 0.996            & 0.996            & 0.997           & 0.997           \\ \midrule
		$\sigma_\mathrm{hf}^*$ [neV] & 31.1       & 12.1   & 14.7             & 16.0             & 17.5            & 20.0            \\ \midrule
		$n^*$                        & 521        & 12.5   & 21.6             & 32.37            & 37.8            & 45.8            \\ \midrule
	\end{tabular}
	\caption{Fitting parameters (see Eqs.~(\ref{eq:dd_hf}, \ref{eq:dd_n})) for the sequential, direct, three- and five-steps protocols versus the noise strength and number of shuttles.
	The protocols in which dynamical decoupling is applied are denoted by (DD), other protocols do not include quantum gates during the shuttling.}
	\label{tab:dd_parameters}
\end{table*}

In practice, the successful implementation of this protocol in an experimental device requires a fast switch of the SOI.
However, this can be challenging on some devices.
A workaround is to take advantage of the fact that the rotation angles depend not only on the value of $x_\mathrm{SOI}$ but also on the number of sites.
Setting a specific value of $x_\mathrm{SOI}$ and determining the required QDA length, we can effectively apply the dynamical decoupling sequence.
For example, with $x_\mathrm{SOI} \sim 0.414$, we can divide the total QDA into three sections consisting of three, five, and three QDs, respectively.
In this configuration, a rotation of $R_Y(\pi/2)$ is applied during transfer in sections with three QDs, while the intermediate section with five QDs is subject to a quantum gate of $R_Y(\pi)$, effectively implementing the Hahn echo protocol.

Furthermore, due to the ability to perform rotations in different axes, such as the $x$- and $y$-axes, these ideas can be extended to more complex dynamical decoupling protocols, such as the Carr-Purcell-Meiboom-Gill (CPMG) scheme \cite{Carr1954, Meiboom1958}.
To implement the CPMG scheme, the QDA must be divided into a minimum of four subarrays, with rotations following the sequence $R_X(\pi/2), R_Y(\pi), R_Y(\pi), R_X(\pi/2)$.
Additionally, applying simultaneous quantum gates during the long-range transfer can facilitate periodic repetitions of high-order dynamical decoupling sequences, leading to stroboscopic saturation.
However, implementing these advanced schemes falls beyond the scope of the present work.

\section{Multiparticle systems}\label{sec:multiparticle_systems}
Our flying qubit architecture works not only for a single particle but also for interacting system, where the Coulomb repulsion between particles plays a crucial role.
Here we present two different scenarios in which the presence of SOI is crucial to obtain the desired results.

\subsection{Quantum state distribution}\label{sec:quantum_state_distribution}
\begin{figure}[!t]
	\centering
	\includegraphics[width=\linewidth]{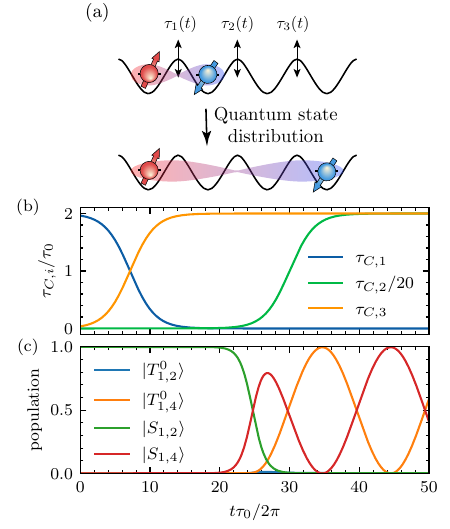}
	\caption{
		(a) Schematic picture of the state distribution of an entangled pair of spin qubits across a quadruple QD, by adiabatically modifying the tunneling barriers $\tau_i(t)$.
		(b) Pulses used for the state distribution, note that the intermediate tunneling rate $\tau_{C, 2}$ is downscaled by a factor of 20.
		(c) Dynamics of a quadruple QD populated with two particles, initialized in the singlet state, after applying the pulses shown in (b).
		The parameters are: $x_\mathrm{SOI}=1$, $T=50\pi/\tau_0$ $a=20$, $b=T/7$, $b_2=3T/5$, $c=T/14$, $E_Z=600\tau_0$, and $U=2500\tau_0$.
	}
	\label{fig:2HH_dynamics}
\end{figure}

In this scenario, we aim to transfer quantum information encoded in the spin of two entangled spins.
We initialize the system in a maximally entangled state, either a singlet or triplet state, which can be represented as $\ket{T_0/S(1, 1)}\equiv(\ket{\uparrow, \downarrow}\pm\ket{\downarrow,\uparrow}) / \sqrt{2}$.
Subsequently, we separate the entangled pair into distant sites of the QDA by tuning the tunneling rates.
The schematic illustration of this idea can be found in Fig.~\ref{fig:2HH_dynamics}~(a).

The minimal system for investigating long-range quantum state distribution with two particles, in terms of DSs, consists of a quadruple QDA \cite{Ban2019}.
Initially, the entangled pair is located at the left-most QD and its neighbor.
We consider the singlet state $\ket{\Psi(0)}=\ket{S_{12}}$, with the particles located at dots 1 and 2.
Similar results can be obtained when the entangled pair is initialized in the unpolarized triplet state.

Inspired by similar protocols in Ref.~\cite{Ban2019}, we use specific pulse shapes:
\begin{equation}
	\begin{split}
		\tau_{C, 1(3)} & = \mp\tau_0\left[\tanh(\frac{t-b}{c}) \mp 1\right], \\
		\tau_{C, 2}    & = a\tau_0\left[\tanh(\frac{t-b_2}{c}) + 1\right].
	\end{split}
\end{equation}
Here, $a=20$, $b=T/7$, $b_2=3T/5$, and $c=T/14$.
Recall that $T$ is the total protocol time.
These pulse shapes are depicted in Fig.~\ref{fig:2HH_dynamics}~(b).
The free parameters ($a$, $b$, and $c$) can be further fine-tuned for improved transfer robustness, although detailed parameter optimization is not within the scope of this work.

The results obtained using these pulses are displayed in Fig.~\ref{fig:2HH_dynamics}~(c).
Notably, the intermediate dot (labeled as $3$) remains unpopulated during the transfer.
As the time reaches $t\sim T/2$, singlet and triplet states distributed at the ends of the QDA begin to oscillate.
The coupling between these states is due to the nonzero SOI.
By selecting the total transfer time, a long-range distribution that preserves spin can be achieved at $t\tau_0/2\pi\sim 45$.
On the other hand, if the transfer protocol ends at $t\tau_0/2\pi\sim 35$, the final state is the triplet state.
In the latter case, the quantum gate implemented simultaneously to the transfer effectively acts as a $Z$ gate applied to one of the two entangled qubits.

\begin{figure}[!t]
	\centering
	\includegraphics[width=\linewidth]{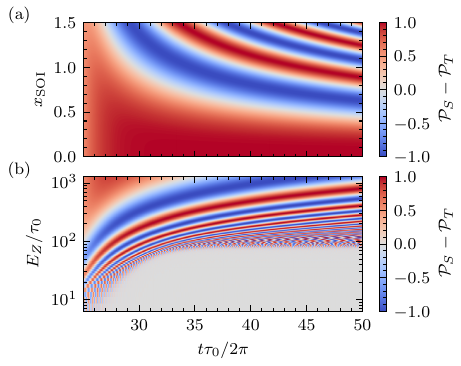}
	\caption{
		(a) Spin polarization $\mathcal{P}_S-\mathcal{P}_T$ in quadruple QDA after a state distribution, versus the protocol time and $x_\mathrm{SOI}$.
		(b) Spin polarization versus the protocol time and Zeeman splitting $E_Z$.
		Other parameters are the same as those used in Fig.~\ref{fig:2HH_dynamics}.
	}
	\label{fig:2HH_polarization}
\end{figure}

For a more detailed analysis of the quantum state distribution, we define the final spin polarization at a given time during the transfer protocol, as follows
\begin{equation}
	\mathcal{P}_{S(T)}(t) \equiv \abs{\braket{S(T^0)_{1,4}}{\Psi(t)}}^2.
\end{equation}
Fig.~\ref{fig:2HH_polarization}~(a) illustrates that, for negligible spin-flip tunneling rate (i.e., $x_\mathrm{SOI}\rightarrow0$), the distributed pair remains in the singlet state.
However, as we increase the SOI to $x_\mathrm{SOI}\geq 0.5$, the system starts to oscillate between the singlet and triplet states.
Higher values of $x_\mathrm{SOI}$ result in a higher oscillation frequency, enabling high-fidelity state distribution plus a change in the spin polarization at shorter total times.

The final spin polarization can also be controlled by adjusting the magnetic field applied to the sample, as shown in Fig.~\ref{fig:2HH_polarization}~(b).
If the magnetic field is not high enough, the polarized triplet states ($T_\pm$) are close in energy to the singlet and the unpolarized triplet states.
The coupling to polarized triplet states acts as a leakage out of the computational basis.
Thus, the high-fidelity transfer can only be achieved if $E_Z\gg \tau_0$.

\;

\subsection{Half-filling regime}\label{sec:half_filling}
\begin{figure}[!t]
	\centering
	\includegraphics[width=\linewidth]{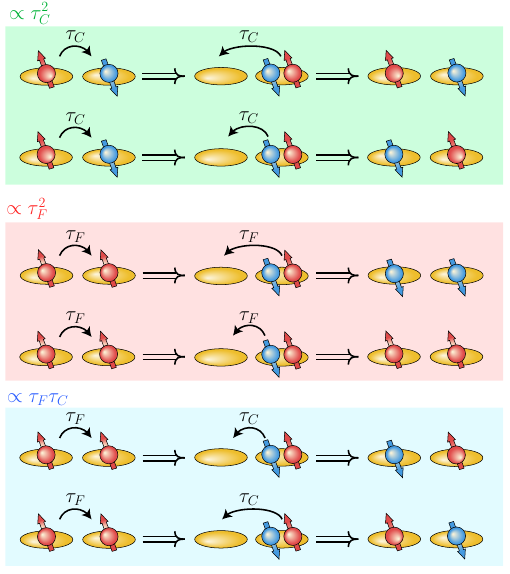}
	\caption{
		Different possible paths for second-order tunneling in the half-filling regime combining spin-conserving and spin-flip tunneling rates.
		There exist paths that are second order in the spin-conserving tunneling rate (top, green), second order in the spin-flip tunneling rate (center, pink), and a combination between spin-conserving and spin-flip tunneling rate (bottom, blue).}
	\label{fig:half_filling_paths}
\end{figure}

Another situation in which the presence of SOI is advantageous for the application of long-range transfer is a QDA in the half-filling regime.
This situation involves a system of $N$ sites with a total of $N$ particles, where all QDs are in resonance and a magnetic field is applied perpendicular to the QD plane.
In the limit of a large Coulomb interaction, where $U$ is much larger than the other energy scales of our problem, we employ a Schrieffer-Wolff transformation to trace out the double-occupied states.
In Fig.~\ref{fig:half_filling_paths}, we show the different possible paths for second-order tunneling processes in the half-filling regime.
If the applied magnetic field is sufficiently strong $E_Z \gg \tau_{C, i},\tau_{F, i}$, the subspaces with fixed total spin projections are separated by a considerable energy gap.
In this case, the coupling between states is primarily governed by the spin-conserving tunneling rates, while the spin-flip tunneling rates renormalize the energies of each level.
For more details on the effective exchange model, see Appendix~\ref{app:hf_hamiltonian}.

The matrix representation of the effective Hamiltonian within the $S_z=+1/2$ subspace, written on the basis of $(\ket{\downarrow, \uparrow, \uparrow}, \ket{\uparrow, \downarrow, \uparrow}, \ket{\uparrow, \uparrow, \downarrow})$, can be expressed as
\begin{widetext}
	\begin{equation}
		H_\text{eff} =\mqty(
		-J^{CC}_1-J^{FF}_2 & J^{CC}_1 & 0 \cr
		J_1^{CC} & -J_1^{CC}-J_2^{CC} & J_2^{CC} \cr
		0 & J_2^{CC} & -J_2^{CC}-J_{1}^{FF} \cr).
		\label{eq:Hamiltonian_eff_TQD_2HH}
	\end{equation}
\end{widetext}

Here, $J^{ab}_i$ represents the exchange coupling between the dots $i$-th and $(i+1)$-th, with $J^{ab}_i \equiv \tau_{a,i} \tau_{b,i}/U$, where $a,b=\{C, F\}$ represents the spin-conserving and spin-flip tunneling rates, respectively.
A common factor $E_Z$ has been subtracted from the diagonal elements of the Hamiltonian.
We will mainly focus on the $S_z=+1/2$ subspace, although analogous results can be applied to the $S_z = -1/2$ subspace.
The effective Hamiltonian resembles a $\Lambda$ system, with a central state ($\ket{\uparrow, \downarrow, \uparrow}$) coupled to the other two states.
To simplify the notation, we define $\ket{i}$ as the state in which all dots are populated with a single spin-up particle, except for the $i$-th site where there is a spin-down particle.

For the achievement of long-range spin transfer, it is imperative to have a DS, an instantaneous eigenstate that exclusively has weight in the initial and final desired states.
A condition for the existence of this DS in a $\Lambda$ system is that $E_1 = E_3$, where $E_i \equiv \bra{i}H_\text{eff}\ket{i}$ \cite{Vitanov2017}.
In the absence of SOI, this condition is satisfied only if $J_1^{CC} = J_2^{CC}$ at all times.
However, when this condition is imposed, the DS remains constant at $\ket{\text{DS}} = (\ket{1}+\ket{3})/\sqrt{2}$ throughout the process, without dependence on the tunneling rates, making the spin transfer unachievable.

To enable long-range spin transfer without the presence of SOI, it is necessary to apply magnetic field gradients in combination with AC driving on the tunneling barriers, wherein the pulse envelope undergoes slow changes \cite{Gullans2020}.
Nonetheless, generating large magnetic field gradients could pose challenges in experimental setups, and AC fields have the potential to induce heating in the sample, leading to significant transfer errors.
These challenges can be avoided by considering holes that exhibit nonzero SOI.

Setting $\tau_{F,i}/\tau_{C, i} = 1$, i.e., $J_i^{CC} = J_{i}^{FF}=J_i$, the Hamiltonian in Eq.~(\ref{eq:Hamiltonian_eff_TQD_2HH}), after subtracting a common factor $-(J_1+J_2)$ in the diagonal elements, reads
\begin{equation}
	H_\text{eff} =\mqty(
	0 & J_1 & 0 \cr
	J_1 & 0 & J_2 \cr
	0 & J_2 & 0 \cr).
    \label{eq:TQD_Hamiltonian}
\end{equation}

With this effective Hamiltonian, we perform an STA protocol to transfer the spin qubit.
For this purpose, we consider an ansatz for the wave function similar to the one presented in Eq.~(\ref{eq:ansatz}) to determine time-dependent pulses $J_1$ and $J_2$ to achieve long-range spin transfer of a single spin in a TQD containing three particles.
The system is initially prepared in the state $\ket{1}$.

We consider the DS obtained from the effective Hamiltonian in Eq.~(\ref{eq:TQD_Hamiltonian}):
\begin{equation}
	\ket{\mathrm{DS}}=\sin\theta \ket{1}-\cos\theta\ket{3},
\end{equation}
where $\tan\theta \equiv J_2/J_1$, and perform the STA protocol to transfer the spin down between the outer dots of the QDA.

It is interesting to note that the effective model is the same as the one obtained for a single particle with renormalized second-order hopping.
That means that the discussion regarding the different protocols and the effect of noise for the single particle in the QDA is equivalent and can be extended to the half-filling regime.

\section{Conclusions}\label{sec:conclusions}
In this work, we investigate the control of flying spin qubits in semiconductor quantum dot arrays in the presence of spin-orbit interaction (SOI).
We demonstrate that the transfer fidelity can be significantly enhanced by engineering the tunneling rates of the barriers appropriately.
A detailed analysis comparing various protocols reveals that Shortcuts to Adiabaticity (STA) emerges as the optimal protocol in terms of speed and minimal population of the middle dots, ensuring rapid and high-fidelity transference even in the presence of dephasing.
Importantly, STA is the highest fidelity protocol for large quantum dot arrays, being therefore more suitable for scalability purposes.

We introduce a new framework to control and tune the spin rotations induced by SOI during the transfer which allows simultaneously the implementation of one-qubit quantum gates.
Furthermore, we propose how to implement dynamical decoupling protocols to mitigate the effects of hyperfine interaction and enhance transfer fidelity.
The application of simultaneous one-qubit gates during the transfer is advantageous to generate entanglement between two distant nodes.
Expanding our analysis to multiparticle systems, we demonstrate that the presence of SOI facilitates long-range entanglement distribution across the quantum dot array and enables long-range spin transfer in the half-filling regime.

All the results shown here are highly dependent on the existence of SOI, which plays a crucial role in the control and manipulation of the flying spin qubit.
The presence of strong SOI is responsible for the emergence of dark states with a mixture of spin states, which are essential for the implementation of quantum gates during the shuttling.
Furthermore, for the multiparticle systems studied here, the long-range transfer of quantum information is only possible due to the presence of SOI.

The generality of our results extends to both holes and electron spin qubits in the presence of SOI, hosted in different semiconductor materials such as germanium or silicon, nanowires, or FinFET devices.
In summary, our findings not only advance the theoretical understanding of quantum systems with SOI but also pave the way for practical applications.
They underscore the advantageous role of SOI to implement long-range quantum information transfer protocols in quantum dot arrays, and show their feasibility as quantum links between processors, enhancing the quantum chip connectivity.
Furthermore, the integration of simultaneous quantum gates into transfer protocols offers the potential for significant acceleration of specific quantum information procedures.

\section*{Acknowledgments}
G.P. and D.F.F. are supported by Spain's MINECO through Grant No. PID2020-117787GB-I00 and by the CSIC Research Platform PTI-001.
G.P. and D.F.F. also acknowledge the agreement between Carlos III University and the CSIC through the UA.
D.F.F. acknowledges support from FPU Program No. FPU20/04762.
Y.B. acknowledges support from the Spanish Government via the project PID2021-126694NA-C22 (MCIU/AEI/FEDER, EU).

\appendix

\section{Effective spin-flip tunneling rate} \label{app:effective_spin_flip}
In germanium or silicon, the Dresselhaus SOI vanishes because the system does not present bulk inversion asymmetry (BIA).
Rashba interaction, due to structure inversion asymmetry (SIA), is the main SOI mechanism present in the material.
Furthermore, depending on the sample configuration, the dominant term for Rashba SOI is linear or cubic in momentum \cite{Rashba1988, Luo2010, Szumniak2012, Liu2022}.
Finally, this discussion can be extended to other semiconductor materials, such as GaAs, after the inclusion of BIA.

The model for a double QD in which a single particle is confined along the out-of-plane ($z$) axis \cite{Altarelli1988, Platero1989} can be described by a quartic harmonic confining potential with dot separation $2a$.
Additionally, there is an out-of-plane magnetic field $B$ resulting in a Zeeman splitting for the spin degree of freedom.
The Hamiltonian reads as follows \cite{Luttinger1955, Winkler2003}
\begin{equation}
	\begin{split}
		H_0 =& \frac{\pi_x^2+\pi_y^2}{2m^*} + \frac{1}{2}m^*\omega_0^2\left(\frac{(x^2-a^2)^2}{4a^2}+y^2\right)\\
		&+\frac{g}{2}\mu_B B\sigma_z,
	\end{split}
	\label{eq:H_0}
\end{equation}
where $\boldsymbol{\pi}=\mathbf{p}+e\mathbf{A}$ is the canonical momentum and $\sigma_i$ the Pauli matrices.
The total Hamiltonian, taking into account Rashba ($\alpha$) and Dresselhaus ($\beta$) SOI, both with linear and cubic terms, is described by $H = H_0+H_\text{SOI}$, where $H_\text{SOI} \equiv H_\alpha^{(1)} + H_\alpha^{(3)}+H_\beta^{(1)}+H_\beta^{(3)}$ with
\begin{equation}
	\begin{split}
		H_\alpha^{(1)} = & i\alpha^{(1)}(\sigma_-\pi_+ - \sigma_+\pi_-),                   \\
		H_\alpha^{(3)} = & i\alpha^{(3)}(\sigma_+\pi_-^3-\sigma_- \pi_+^3),                \\
		H_\beta^{(1)} =  & \beta^{(1)}(\sigma_+\pi_+ + \sigma_-\pi_-),                     \\
		H_\beta^{(3)} =  & \beta^{(3)}(\sigma_+\pi_-\pi_+\pi_- + \sigma_-\pi_+\pi_-\pi_+).
	\end{split}
\end{equation}
The Hamiltonian in Eq.~(\ref{eq:H_0}) can be exactly solved by obtaining the Fock-Darwin states $\psi_{nl}^\text{FD}(x, y)$.
Considering just a single orbital per dot, we restrict ourselves to $\psi_{00}^\text{FD}(x, y) = \frac{1}{\sqrt{\pi}b}\exp{-(x^2+y^2)/2b^2}$, where we define $b^2\equiv \hbar/m^*\sqrt{\omega_0^2+\omega_L^2}$, with the Larmor frequency $\omega_L=eB/2m^*$.
The magnetic vector potential is defined in the symmetric gauge $\mathbf{A} = B(-y, x, 0)/2$.
Transforming the Fock-Darwin states under this gauge, they acquire a phase as
\begin{equation}
	\psi_{00}^{L/R} = e^{\pm iya/2l_B^2}\psi_{00}^\text{FD}(x\pm a, y).
\end{equation}

The finite overlap between the left and right wave functions is given by
\begin{equation}
	S=\braket{\psi_{00}^L}{\psi_{00}^R} = \exp(-\frac{a^2m^*(\omega_0^2 + 2\omega_L^2)}{\hbar\omega}),
\end{equation}
with $\omega\equiv\sqrt{\omega_0^2+\omega_L^2}$.

We orthogonalize the Wannier states as $\ket{L/R}=\sqrt{N}(\psi_{00}^{L/R}-\gamma\psi_{00}^{R/L})$, where the normalization is given by $N = (1-2\gamma S+\gamma^2)^{-1}$ and $\gamma=(1-\sqrt{1-S^2})/S$.
The spin-flip terms for the SOI Hamiltonian are given by:
\begin{eqnarray}
	\bra{L\downarrow}H_\alpha^{(1)}\ket{R\uparrow} & = -\alpha^{(1)}N(1-\gamma^2)\frac{am^*\omega_0^2}{\omega}S, \nonumber\\
	\bra{R\downarrow}H_\alpha^{(1)}\ket{L\uparrow} & = \alpha^{(1)}N(1-\gamma^2)\frac{am^*\omega_0^2}{\omega}S, \nonumber \\
\label{eq:H_alpha_1}
\end{eqnarray}
for the linear-in-momentum Rashba term, and

\begin{eqnarray}
	\bra{L\downarrow}H_\alpha^{(3)}\ket{R\uparrow} & = -\alpha^{(3)}N(1-\gamma^2)\left(\frac{am^*\omega_0^2}{\omega}\right)^3S, \nonumber\\
	\bra{R\downarrow}H_\alpha^{(3)}\ket{L\uparrow} & = \alpha^{(3)}N(1-\gamma^2)\left(\frac{am^*\omega_0^2}{\omega}\right)^3S,  \nonumber\\
\label{eq:H_alpha_3}
\end{eqnarray}
for the cubic-in-momentum Rashba term.
The corresponding matrix elements for the linear- and cubic-in-momentum Dresselhaus SOI are given by
\begin{eqnarray}
	\bra{L\downarrow}H_\beta^{(1)}\ket{R\uparrow} & =i\beta^{(1)}N(1-\gamma^2)\frac{am^*\omega_0^2}{\omega}S,  \nonumber \\
	\bra{R\downarrow}H_\beta^{(1)}\ket{L\uparrow} & =-i\beta^{(1)}N(1-\gamma^2)\frac{am^*\omega_0^2}{\omega}S, \nonumber \\
\label{eq:H_beta_1}
\end{eqnarray}

\begin{widetext}
	\begin{equation}
		\begin{split}
			\bra{L\downarrow}H_\beta^{(3)}\ket{R\uparrow} & =i\beta^{(3)}N(1-\gamma^2)\frac{a{m^*}^2\omega_0^2}{\omega^3}(a^2m^*\omega_0^4-2\hbar\omega(\omega_0^2+2\omega_L^2))S,  \\
			\bra{R\downarrow}H_\beta^{(3)}\ket{L\uparrow} & =-i\beta^{(3)}N(1-\gamma^2)\frac{a{m^*}^2\omega_0^2}{\omega^3}(a^2m^*\omega_0^4-2\hbar\omega(\omega_0^2+2\omega_L^2))S.\\
		\end{split}
		\label{eq:H_beta_3}
	\end{equation}
\end{widetext}
	
We compute $\bra{L\downarrow}H_\text{SOI}\ket{R\uparrow}$ and $\bra{R\downarrow}H_\text{SOI}\ket{L\uparrow}$, while the other two matrix elements can be obtained by imposing the hermiticity of the resulting Hamiltonian.
From Eqs.~(\ref{eq:H_alpha_1}-\ref{eq:H_beta_3}) we obtain that by including the linear or cubic term, both in the Rashba and Dresselhaus SOI terms, the phenomenology of the model remains the same, and the spin-flip tunneling rate can be written as

\begin{equation}
	-\tau_F \equiv \bra{R\downarrow}H_\mathrm{SOI}\ket{L\uparrow}=-\bra{L\downarrow}H_\mathrm{SOI}\ket{R\uparrow}.
\end{equation}

\section{Effective Model for a QDA with \texorpdfstring{$N>3$}{N>3}}\label{app:larger_arrays}
\begin{figure*}[!t]
	\centering
	\includegraphics[width=\linewidth]{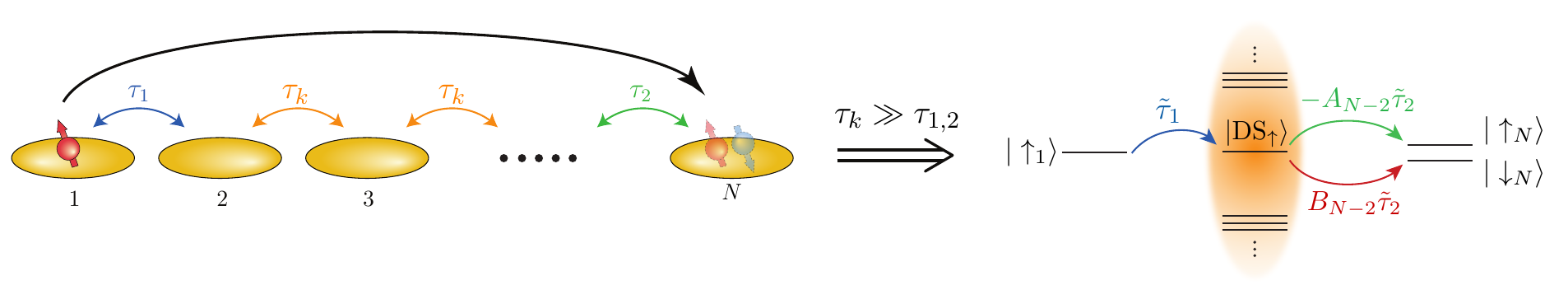}
	\caption{Scheme of a linear N QDA, and straddling pulses $\tau_k$ applied to the intermediate dots.
	The system is initialized with a spin-up particle in the leftmost QD.
	Depending on the number of sites and the effective SOI, the transfer protocol conserves or inverts the spin of the particle.
	If the straddling pulse is much larger than the boundary pulses $\tau_1$ and $\tau_2$, the dots in the bulk form a dressed state connecting the initial state with the final one.
	The effective coupling rate between the dressed and the final states depends on the functions $A_k$ and $B_k$, which are related to the number of sites and the SOI ratio.}
	\label{fig:effective_long_array}
\end{figure*}

In this section, we obtain an effective model for a linear QDA with more than three sites, populated with a single particle.
To transfer the particle between the ends of the chain, we use straddling pulses, with all intermediate tunneling rates $\tau_{C, k}$, with $1<k<N-1$, equal to each other, see Fig.~\ref{fig:effective_long_array}.
Setting $\tau_{F, k}=x_\mathrm{SOI}\tau_{C, k}$, and after decoupling the leftmost and rightmost dots from the rest of the array $\tau_{C, 1} = \tau_{C, N-1} = 0$, we obtain two DSs described by
\begin{eqnarray}
	\ket{\text{DS}_\uparrow} = \frac{1}{\sqrt{(N-1)/2}}\sum_{k=1}^{(N-1)/2}&\left(A_{2k-1}\ket{\uparrow_{2k}}\nonumber \right.\\
	&\left.+B_{2k-1}\ket{\downarrow_{2k}}\right), \nonumber\\
	\label{eq:DS_eff_1}
\end{eqnarray}

\begin{eqnarray}
	\ket{\text{DS}_\downarrow} = \frac{1}{\sqrt{(N-1)/2}}\sum_{k=1}^{(N-1)/2}&\left(B_{2k-1}\ket{\uparrow_{2k}}\right. \nonumber\\
	&\left.+A_{2k-1}\ket{\downarrow_{2k}}\right).\nonumber\\
	\label{eq:DS_eff_2}
\end{eqnarray}

Here, we have defined the functions
\begin{eqnarray}
	A_k & = (-1)^{(k+1)/2}\cos\left[\arctan(x_\mathrm{SOI})(1-k)\right], \nonumber \\
	B_k & = (-1)^{(k+1)/2}\sin\left[\arctan(x_\mathrm{SOI})(1-k)\right].\nonumber \\
\end{eqnarray}

In the limit of zero SOI, i.e., $x_\mathrm{SOI}=0$, the above expressions simplify as $A_i = (-1)^{(i+1)/2}$ and $B_i=0$, obtaining the same DSs as those presented in \cite{Gullans2020}, which only contain states with the same spin projection.

The DSs shown in Eqs.~(\ref{eq:DS_eff_1}), (\ref{eq:DS_eff_2}) represent dressed states formed by all the dots in the bulk of the QDA.
We couple the edge QDs by setting $\tau_{C, 1}, \tau_{C, 2}\ll \tau_{C, k}$, which allows us to reduce the large QDA to an effective six-level system on the basis $\left\{\ket{\uparrow_1}, \ket{\downarrow_1},\ket{\text{DS}_\uparrow}, \ket{\text{DS}_\downarrow}, \ket{\uparrow_N}, \ket{\downarrow_N}\right\}$

\begin{widetext}
	\begin{equation}
		H =\mqty(
		0 & 0 & \tilde{\tau}_1(t) & 0 & 0 & 0 \cr
		0 & 0 & 0 & \tilde{\tau}_1(t) & 0 & 0 \cr
		\tilde{\tau}_1(t) & 0 & 0 & 0 & -A_{N-2}\tilde{\tau}_2(t) & B_{N-2}\tilde{\tau}_2(t) \cr
		0 & \tilde{\tau}_1(t) & 0 & 0 & B_{N-2}\tilde{\tau}_2(t) & A_{N-2}\tilde{\tau}_2(t) \cr
		0 & 0 & -A_{N-2}\tilde{\tau}_2(t) & B_{N-2}\tilde{\tau}_2(t) & 0 & 0 \cr
		0 & 0 & B_{N-2}\tilde{\tau}_2(t) & A_{N-2}\tilde{\tau}_2(t) & 0 & 0 \cr),
		\label{eq:H_eff}
	\end{equation}
\end{widetext}
with $\tilde{\tau}_{1(2)}(t) = \tau_{C, 1(2)}(t)\sqrt{2/(N-1)}$.
In the above equation, we have already used the fact that $A_1 = -1$ and $B_1 =0$.
This system can be identified with a QDA of three sites under transformation $\ket{DS_{\uparrow(\downarrow)}}\rightarrow\ket{\uparrow(\downarrow)_2}$.
The final spin projection of the transferred particle depends on the number of sites in the total QDA and the value of the effective SOI.
However, a long-range transfer with a minimal population in the middle sites is possible even in large QDAs in the presence of SOI.

\section{Dark states for out-of-resonance conditions}\label{app:dark_states_in_real}
Due to systematic errors in the experimental implementation of virtual gates, the energy levels of the system can be shifted, leading to a deviation from the ideal Hamiltonian.
The most crucial requirement for the existence of DSs is that the energy levels of the initial and final states are degenerate.
We compute the fidelity of spin qubit transfer in the presence of randomly time-constant and uncorrelated errors in the center dots, i.e., $\varepsilon_i \rightarrow \varepsilon_i + \delta_i$ for $2\leq i \leq N - 1$, where $\delta_i$ is a random number uniformly distributed in the interval $[-\delta_\varepsilon, \delta_\varepsilon]$.

\begin{figure}[!t]
	\centering
	\includegraphics[width=\linewidth]{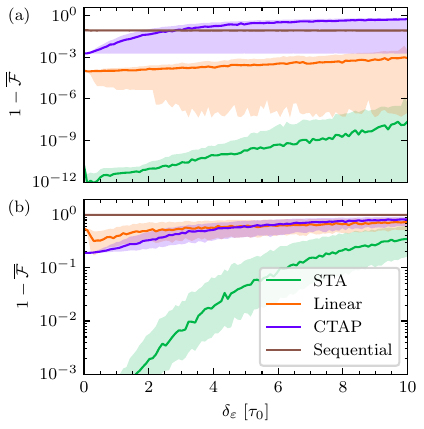}
	\caption{Infidelity of different protocols in (a) $N=3$ and (b) $N=15$ QDA against the fluctuation strength in the energy levels of the intermediate dots.
	Colored solid lines represent the mean infidelity after $10^3$ noise realizations, while the shaded area indicates the interquartile range.
	Each different protocol is represented by a different color: STA (green), linear (orange), CTAP (purple), and sequential (brown).
	Other parameters are $\tau_s=10\tau_0$, $T=100 / \tau_0$, $x_\mathrm{SOI}=1$, $\sigma_\mathrm{CTAP} = \sigma_\mathrm{bulk} = T / 6$.
	For the sequential transfer, $\varepsilon_\mathrm{max}=100\tau_0$ is used.}
	\label{fig:DS_intermediate_fluctuations}
\end{figure}

If Fig.~\ref{fig:DS_intermediate_fluctuations}, we show the infidelity of the different protocols for a QDA with $N=3$ and $N=15$ sites.
Since the distribution over $10^3$ noise realizations is highly skew, we use the interquartile range to represent the spread of the data, as the mean value and the standard deviation are not representative.
We observe that, even in presence of moderate energy level fluctuations, the DS remains present, allowing for long-range transfer with high fidelity compared with the sequential transfer.
Given that $\delta_\varepsilon \ll \varepsilon_\mathrm{max}$, the sequential transfer is almost insensitive to the fluctuations in the energy levels.
However, for large QDAs, the sequential transfer is not feasible due to very low fidelity.
Among the different long-range protocols, STA achieves the highest fidelity, outperforming the other ones.

\section{Charge noise}\label{app:charge_noise}
The success of quantum state transfer through a quantum bus hinges on its resilience against experimental errors throughout the protocol.
In the context of spin qubits in semiconductor QDs, the most significant source of noise arises from electric field fluctuations, often referred to as charge noise.

We model charge noise by considering time-dependent fluctuations in both the energy of each QD and the barriers that define the tunneling rates.
We parameterize the charge noise as \cite{Gullans2019}
\begin{gather}
	\varepsilon_i^n(t) = \varepsilon_i+\delta_\varepsilon\nu_i(t), \\
	\tau_{C,i}^n(t) = \tau_{C,i}(t)+\delta_\tau\tilde{\nu}_i(t).
	\label{eq:tunneling_noise}
\end{gather}
Here, the superscript $n$ indicates the presence of noise.
Furthermore, we assume uncorrelated noise terms, which means there are no spatial correlations between them, with $\expval{\nu_i(t)\nu_j(t)} = \expval{\tilde{\nu}_i(t)\tilde{\nu}_j(t)}= \delta_{ij}$ and $\expval{\tilde{\nu}_i(t)\nu_j(t)}=0$.
To model charge noise, we employ a characteristic spectral density of pink noise, given by $S(f)\propto1/f$ for frequencies within the range $f_\mathrm{min}<f<f_\mathrm{max}$.
For frequencies below $f_\mathrm{min}$, we assume white noise without frequency dependence, and for frequencies above $f_\mathrm{max}$, we adopt Brownian noise with $S(f)\propto 1/f^2$.
The parameters $\delta_\varepsilon$ and $\delta_\tau$ represent the strength of the noise for detuning and tunneling rates, respectively.
The spin-flip tunneling rates in the presence of noise are defined by the relation $\tau_{F, i}^n(t) = x_\mathrm{SOI}\tau_{C, i}^n(t)$.

For simulating transfer under the influence of charge noise, the system is initialized at $\ket{\Psi(0)}=\ket{\uparrow_1}$.
We then average the solution of the time-dependent Schrödinger equation over $10^3$ independent noise realizations.
\begin{figure}[!t]
	\centering
	\includegraphics[width=\linewidth]{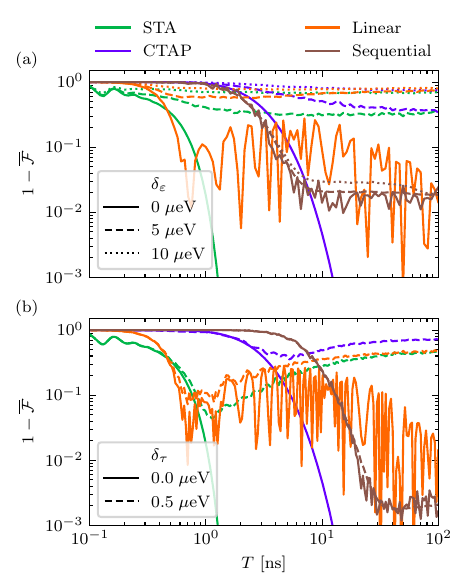}
	\caption{
		Average infidelity of a long-range transfer in a seven QDA versus the total time of the protocol and (a) noise in the detuning $\delta_\varepsilon$, (b) noise in the tunneling rates $\delta_\tau$.
		Each different protocol is represented with a color: STA (green), linear (orange), sequential (brown), and CTAP (purple).
		Solid lines represent zero noise, and dashed lines represent the infidelity in the presence of noise.
		$\tau_0 = 10\;\mu$eV, $\tau_s = 10 \tau_0$, $\varepsilon_\mathrm{max} = 500\; \mu$eV, $f_\mathrm{min}=0.16$ mHz and $f_\mathrm{max}=100$ kHz.}
	\label{fig:noisy_transfer}
\end{figure}
When charge noise is present due to fluctuations in the energy levels, Fig.~\ref{fig:noisy_transfer} (a), the highest fidelity for a short time is obtained with STA.
For larger times, CTAP and STA perform similarly to each other, obtaining a transfer fidelity close to $\mathcal{F}\sim 0.99$.
However, the sequential protocol is more robust against charge noise (dashed brown curve), so if the total protocol time is long enough, such that the adiabatic condition is met, the fidelity is close to $\mathcal{F}\sim 0.999$.
On the other hand, the linear ramp shows much worse results than the other ones.
Linear ramps are very sensitive to charge noise in the energy levels and give rise to a maximum fidelity lower than $0.9$.
Both CTAP and STA are much more sensitive to errors in tunneling rates, Fig.~\ref{fig:noisy_transfer}~(b), acquiring fidelities lower than in the previous case.
However, the sequential protocol is still robust against this error, obtaining fidelities close to the noiseless case with long protocol times.

\begin{figure*}[thb!]
	\centering
	\includegraphics[width=\linewidth]{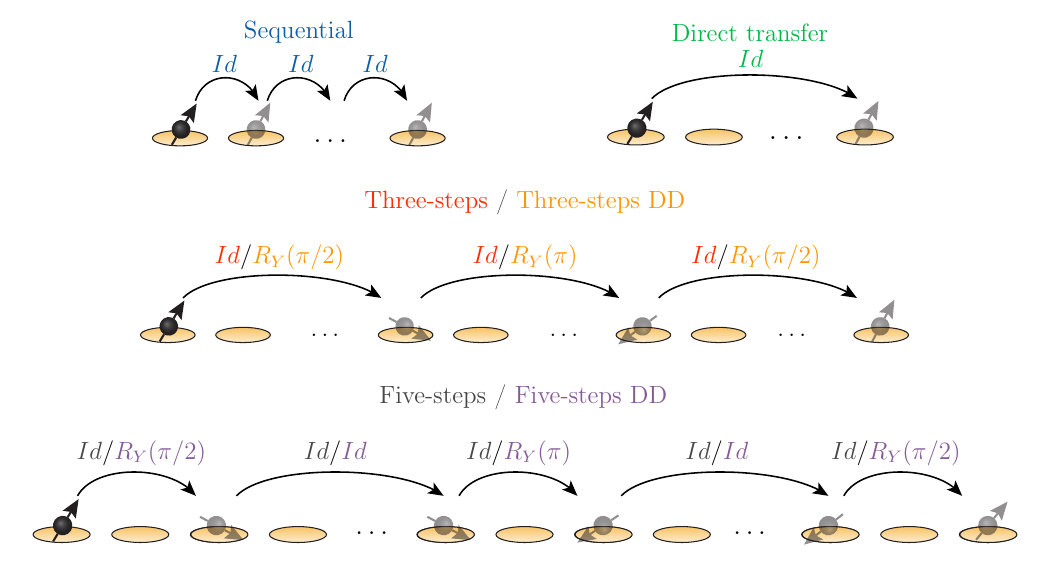}
	\caption{
		Schematics of different transfer protocols.
		In the top row, sequential (blue) and direct transfer (green) protocols are depicted, which keep the particle's spin projection constant during the transfer, denoted by the identity gate (Id).
		In the middle row, we show two protocols that divide the linear array into three subarrays.
		During the three-steps protocol (red), no quantum gates are applied in each long-range transfer.
		On the other hand, for the three-steps DD protocol (orange), the quantum gates $R_Y(\pi/2)$ and $R_Y(\pi)$ are applied during the transfer.
		Finally, in the bottom row, we show the five-steps (gray) and five-steps DD (violet) protocols, in which the linear array is divided into five subarrays.
		}
	\label{fig:dynamical_decoupling_schematics}
\end{figure*}

In a more realistic case, more noise sources affect the system.
Therefore, to improve the transfer fidelity, it is mandatory to perform the transfer as fast as possible.
Furthermore, it is also beneficial for quantum algorithms to speed up the transfer between computing nodes so that the total algorithm is performed in a reasonable time.
A more detailed analysis of the different noise sources in a realistic set-up must be done to determine which protocol is more desirable.

\section{Dynamical decoupling schemes}\label{app:dynamical_decoupling}

In this section, we present different protocols used for the DD schemes discussed in the main text.
The schematics for the protocols are shown in Fig.~\ref{fig:dynamical_decoupling_schematics}.
In the sequential protocol, the spin qubit is transported through all dots, and no quantum gate is performed during the transfer.
On the contrary, the direct transfer protocol implies that the middle dots are not populated, but DD is not applied.
On the other hand, when the linear array is divided into three subarrays, we can perform different quantum gates during the transfer.
In the first and third subarrays, we perform an $R_Y(\pi/2)$ rotation, and an $R_Y(\pi)$ rotation in the second one, performing a DD protocol simultaneously with the particle transfer.
We can also divide the total array into five subarrays, so the DD is applied during the first, third, and fifth subarrays.
We benchmark the results of the DD schemes by performing a similar transfer in the absence of quantum gates applied during the transfer.

\section{Half-filling regime Hamiltonian}\label{app:hf_hamiltonian}
In this section, we will derive the effective Hamiltonian for a linear QDA in the half-filling regime with SOI, in the limit of high Coulomb interaction.
To address this scenario, through a Schriffer-Wolf transformation, we trace out the double-occupied states, resulting in an effective exchange Hamiltonian.
We consider a double QD populated with two particles.
In Fig.~\ref{fig:half_filling_paths} of the main text, we illustrate the possible tunneling paths combining spin-conserving and spin-flip tunneling rates that a particle can perform in a DQD.
The complete set of elements can be derived through spin inversion.

The original Hamiltonian $H_0$ is described in Eq.~(\ref{eq:original_hamiltonian}) of the main text.
Both the spin-conserving tunneling term in Eq.~(\ref{eq:tau_hamiltonian}), and the effective SOI in Eq.~(\ref{eq:SOI_hamiltonian}), are treated as perturbations to the original Hamiltonian via $V=H_\tau + H_\mathrm{SOI}$.
We perform a Schrieffer-Wolff transformation at first order, by solving: $V + \left[S, H_0\right] = 0$.
In the next step, we simplify the effective model by considering $U \gg E_Z, \tau_{C, i}, \tau_{F, i}$.
Finally, the effective Hamiltonian, for $\varepsilon_i = 0$, given by $H'=H_0 + \left[S,V\right]/2$, reads
\begin{equation}
	\begin{split}
		H' = & E_Z(\sigma_z^1 + \sigma_z^2)+ 2J^{CF}(\sigma_x^1\sigma_z^2-\sigma_z^1\sigma_x^2) \\
				& + J^{CC}(\sigma_x^1\sigma_x^2+\sigma_y^1\sigma_y^2+\sigma_z^1\sigma_z^2 - 1/4)   \\
				& + J^{FF}(-\sigma_x^1\sigma_x^2+\sigma_y^1\sigma_y^2-\sigma_z^1\sigma_z^2 - 1/4),
		\label{eq:HH_effective_Hamiltonian}
	\end{split}
\end{equation}
where $\sigma_\alpha^i$ corresponds to the Pauli matrix $\alpha=\left\{x,y,z\right\}$ acting at the $i=\left\{1, 2\right\}$ site.
The various exchange couplings are denoted as $J^{ab} \equiv \tau_a \tau_b/U$, where $a,b=\{C, F\}$ represent the spin-conserving and spin-flip tunneling rates, respectively.
The effective model can be extended to longer arrays, with an arbitrary number of sites, as
\begin{equation}
	\begin{split}
		H' = & E_Z\sum_i \sigma_z^i + 2\sum_{<i,j>}J^{CF}_i(\sigma_x^i\sigma_z^j-\sigma_z^i\sigma_x^j)        \\
				& + \sum_{<i,j>}J^{CC}_i(\sigma_x^i\sigma_x^j+\sigma_y^i\sigma_y^j+\sigma_z^i\sigma_z^j - 1/4)   \\
				& + \sum_{<i,j>}J^{FF}_i(-\sigma_x^i\sigma_x^j+\sigma_y^i\sigma_y^j-\sigma_z^i\sigma_z^j - 1/4).
		\label{eq:HH_effective_Hamiltonian_N}
	\end{split}
\end{equation}

\bibliographystyle{quantum.bst}
\bibliography{references}

\begin{thebibliography}{100}

\bibitem{Loss1998}
Daniel Loss and David~P. DiVincenzo.
\newblock ``{Quantum computation with quantum dots}''.
\newblock \href{https://doi.org/10.1103/physreva.57.120}{Physical Review A {\bf 57}, 120--126}~(1998).

\bibitem{Burkard1999}
Guido Burkard, Daniel Loss, and David~P. DiVincenzo.
\newblock ``{Coupled quantum dots as quantum gates}''.
\newblock \href{https://doi.org/10.1103/physrevb.59.2070}{Physical Review B {\bf 59}, 2070--2078}~(1999).

\bibitem{Ciorga2000}
M.~Ciorga, A.~S. Sachrajda, P.~Hawrylak, C.~Gould, P.~Zawadzki, S.~Jullian, Y.~Feng, and Z.~Wasilewski.
\newblock ``Addition spectrum of a lateral dot from coulomb and spin-blockade spectroscopy''.
\newblock \href{https://doi.org/10.1103/physrevb.61.r16315}{Physical Review B {\bf 61}, R16315--R16318}~(2000).

\bibitem{Petta2005}
J.~R. Petta, A.~C. Johnson, J.~M. Taylor, E.~A. Laird, A.~Yacoby, M.~D. Lukin, C.~M. Marcus, M.~P. Hanson, and A.~C. Gossard.
\newblock ``{Coherent Manipulation of Coupled Electron Spins in Semiconductor Quantum Dots}''.
\newblock \href{https://doi.org/10.1126/science.1116955}{Science {\bf 309}, 2180--2184}~(2005).

\bibitem{Nowack2007}
K.~C. Nowack, F.~H.~L. Koppens, Yu.~V. Nazarov, and L.~M.~K. Vandersypen.
\newblock ``{Coherent Control of a Single Electron Spin with Electric Fields}''.
\newblock \href{https://doi.org/10.1126/science.1148092}{Science {\bf 318}, 1430--1433}~(2007).

\bibitem{Foletti2009}
Sandra Foletti, Hendrik Bluhm, Diana Mahalu, Vladimir Umansky, and Amir Yacoby.
\newblock ``{Universal quantum control of two-electron spin quantum bits using dynamic nuclear polarization}''.
\newblock \href{https://doi.org/10.1038/nphys1424}{Nature Physics {\bf 5}, 903--908}~(2009).

\bibitem{Bluhm2010}
Hendrik Bluhm, Sandra Foletti, Izhar Neder, Mark Rudner, Diana Mahalu, Vladimir Umansky, and Amir Yacoby.
\newblock ``{Dephasing time of {GaAs} electron-spin qubits coupled to a nuclear bath exceeding 200 {$\mu$}s}''.
\newblock \href{https://doi.org/10.1038/nphys1856}{Nature Physics {\bf 7}, 109--113}~(2010).

\bibitem{Vidan2004}
A.~Vidan, R.~M. Westervelt, M.~Stopa, M.~Hanson, and A.~C. Gossard.
\newblock ``{Triple quantum dot charging rectifier}''.
\newblock \href{https://doi.org/10.1063/1.1807030}{Applied Physics Letters {\bf 85}, 3602--3604}~(2004).

\bibitem{Gaudreau2006}
L.~Gaudreau, S.~A. Studenikin, A.~S. Sachrajda, P.~Zawadzki, A.~Kam, J.~Lapointe, M.~Korkusinski, and P.~Hawrylak.
\newblock ``{Stability Diagram of a Few-Electron Triple Dot}''.
\newblock \href{https://doi.org/10.1103/physrevlett.97.036807}{Physical Review Letters {\bf 97}, 036807}~(2006).

\bibitem{Schroeer2007}
D.~Schröer, A.~D. Greentree, L.~Gaudreau, K.~Eberl, L.~C.~L. Hollenberg, J.~P. Kotthaus, and S.~Ludwig.
\newblock ``{Electrostatically defined serial triple quantum dot charged with few electrons}''.
\newblock \href{https://doi.org/10.1103/physrevb.76.075306}{Physical Review B {\bf 76}, 075306}~(2007).

\bibitem{Rogge2008}
M.~C. Rogge and R.~J. Haug.
\newblock ``{Two-path transport measurements on a triple quantum dot}''.
\newblock \href{https://doi.org/10.1103/physrevb.77.193306}{Physical Review B {\bf 77}, 193306}~(2008).

\bibitem{Gaudreau2011}
L.~Gaudreau, G.~Granger, A.~Kam, G.~C. Aers, S.~A. Studenikin, P.~Zawadzki, M.~Pioro-Ladri{\`{e}}re, Z.~R. Wasilewski, and A.~S. Sachrajda.
\newblock ``Coherent control of three-spin states in a triple quantum dot''.
\newblock \href{https://doi.org/10.1038/nphys2149}{Nature Physics {\bf 8}, 54--58}~(2011).

\bibitem{Noiri2022}
Akito Noiri, Kenta Takeda, Takashi Nakajima, Takashi Kobayashi, Amir Sammak, Giordano Scappucci, and Seigo Tarucha.
\newblock ``A shuttling-based two-qubit logic gate for linking distant silicon quantum processors''.
\newblock \href{https://doi.org/10.1038/s41467-022-33453-z}{Nature Communications {\bf 13}, 5740}~(2022).

\bibitem{Creffield2002}
C.~E. Creffield and G.~Platero.
\newblock ``Dynamical control of correlated states in a square quantum dot''.
\newblock \href{https://doi.org/10.1103/physrevb.66.235303}{Physical Review B {\bf 66}, 235303}~(2002).

\bibitem{Hensgens2017}
T.~Hensgens, T.~Fujita, L.~Janssen, Xiao Li, C.~J.~Van Diepen, C.~Reichl, W.~Wegscheider, S.~Das Sarma, and L.~M.~K. Vandersypen.
\newblock ``{Quantum simulation of a Fermi{\textendash}Hubbard model using a semiconductor quantum dot array}''.
\newblock \href{https://doi.org/10.1038/nature23022}{Nature {\bf 548}, 70--73}~(2017).

\bibitem{PerezGonzalez2019a}
Beatriz P{\'{e}}rez-Gonz{\'{a}}lez, Miguel Bello, Gloria Platero, and {\'{A}}lvaro G{\'{o}}mez-Le{\'{o}}n.
\newblock ``{Simulation of 1D Topological Phases in Driven Quantum Dot Arrays}''.
\newblock \href{https://doi.org/10.1103/physrevlett.123.126401}{Physical Review Letters {\bf 123}, 126401}~(2019).

\bibitem{Dehollain2020}
J.~P. Dehollain, U.~Mukhopadhyay, V.~P. Michal, Y.~Wang, B.~Wunsch, C.~Reichl, W.~Wegscheider, M.~S. Rudner, E.~Demler, and L.~M.~K. Vandersypen.
\newblock ``{Nagaoka ferromagnetism observed in a quantum dot plaquette}''.
\newblock \href{https://doi.org/10.1038/s41586-020-2051-0}{Nature {\bf 579}, 528--533}~(2020).

\bibitem{Diepen2021}
C.{\hspace{0.167em}}J. van Diepen, T.-K. Hsiao, U.~Mukhopadhyay, C.~Reichl, W.~Wegscheider, and L.{\hspace{0.167em}}M.{\hspace{0.167em}}K. Vandersypen.
\newblock ``{Quantum Simulation of Antiferromagnetic Heisenberg Chain with Gate-Defined Quantum Dots}''.
\newblock \href{https://doi.org/10.1103/physrevx.11.041025}{Physical Review X {\bf 11}, 041025}~(2021).

\bibitem{Xue2022}
Xiao Xue, Maximilian Russ, Nodar Samkharadze, Brennan Undseth, Amir Sammak, Giordano Scappucci, and Lieven M.~K. Vandersypen.
\newblock ``{Quantum logic with spin qubits crossing the surface code threshold}''.
\newblock \href{https://doi.org/10.1038/s41586-021-04273-w}{Nature {\bf 601}, 343--347}~(2022).

\bibitem{Busl2010}
Maria Busl, Rafael S{\'{a}}nchez, and Gloria Platero.
\newblock ``{Control of spin blockade by ac magnetic fields in triple quantum dots}''.
\newblock \href{https://doi.org/10.1103/physrevb.81.121306}{Physical Review B {\bf 81}, 121306(R)}~(2010).

\bibitem{Busl2013}
M.~Busl, G.~Granger, L.~Gaudreau, R.~Sánchez, A.~Kam, M.~Pioro-Ladrière, S.~A. Studenikin, P.~Zawadzki, Z.~R. Wasilewski, A.~S. Sachrajda, and G.~Platero.
\newblock ``Bipolar spin blockade and coherent state superpositions in a triple quantum dot''.
\newblock \href{https://doi.org/10.1038/nnano.2013.7}{Nature Nanotechnology {\bf 8}, 261--265}~(2013).

\bibitem{Sanchez2014}
R.~S{\'{a}}nchez, G.~Granger, L.~Gaudreau, A.~Kam, M.~Pioro-Ladri{\`{e}}re, S.{\hspace{0.167em}}A. Studenikin, P.~Zawadzki, A.{\hspace{0.167em}}S. Sachrajda, and G.~Platero.
\newblock ``Long-range spin transfer in triple quantum dots''.
\newblock \href{https://doi.org/10.1103/physrevlett.112.176803}{Physical Review Letters {\bf 112}, 176803}~(2014).

\bibitem{Sanchez2014a}
Rafael Sánchez, Fernando Gallego-Marcos, and Gloria Platero.
\newblock ``Superexchange blockade in triple quantum dots''.
\newblock \href{https://doi.org/10.1103/physrevb.89.161402}{Physical Review B {\bf 89}, 161402}~(2014).

\bibitem{Braakman2013}
F.~R. Braakman, P.~Barthelemy, C.~Reichl, W.~Wegscheider, and L.~M.~K. Vandersypen.
\newblock ``{Long-distance coherent coupling in a quantum dot array}''.
\newblock \href{https://doi.org/10.1038/nnano.2013.67}{Nature Nanotechnology {\bf 8}, 432--437}~(2013).

\bibitem{GallegoMarcos2015}
Fernando Gallego-Marcos, Rafael S{\'{a}}nchez, and Gloria Platero.
\newblock ``{Photon assisted long-range tunneling}''.
\newblock \href{https://doi.org/10.1063/1.4913834}{Journal of Applied Physics {\bf 117}, 112808}~(2015).

\bibitem{GallegoMarcos2016}
Fernando Gallego-Marcos, Rafael Sánchez, and Gloria Platero.
\newblock ``Coupled landau-zener-stückelberg quantum dot interferometers''.
\newblock \href{https://doi.org/10.1103/physrevb.93.075424}{Physical Review B {\bf 93}, 075424}~(2016).

\bibitem{Stano2015}
Peter Stano, Jelena Klinovaja, Floris~R. Braakman, Lieven M.~K. Vandersypen, and Daniel Loss.
\newblock ``Fast long-distance control of spin qubits by photon-assisted cotunneling''.
\newblock \href{https://doi.org/10.1103/physrevb.92.075302}{Physical Review B {\bf 92}, 075302}~(2015).

\bibitem{PicoCortes2019}
Jordi Pic{\'{o}}-Cort{\'{e}}s, Fernando Gallego-Marcos, and Gloria Platero.
\newblock ``{Direct transfer of two-electron quantum states in ac-driven triple quantum dots}''.
\newblock \href{https://doi.org/10.1103/physrevb.99.155421}{Physical Review B {\bf 99}, 155421}~(2019).

\bibitem{PicoCortes2021}
Jordi Pic{\'{o}}-Cort{\'{e}}s and Gloria Platero.
\newblock ``{Dynamical second-order noise sweetspots in resonantly driven spin qubits}''.
\newblock \href{https://doi.org/10.22331/q-2021-12-23-607}{Quantum {\bf 5}, 607}~(2021).

\bibitem{Nguyen2017}
Thien Nguyen, Charles~D. Hill, Lloyd C.~L. Hollenberg, and Matthew~R. James.
\newblock ``{Fan-out Estimation in Spin-based Quantum Computer Scale-up}''.
\newblock \href{https://doi.org/10.1038/s41598-017-13308-0}{Scientific Reports {\bf 7}, 13386}~(2017).

\bibitem{Pauka2019}
S.~J. Pauka, K.~Das, R.~Kalra, A.~Moini, Y.~Yang, M.~Trainer, A.~Bousquet, C.~Cantaloube, N.~Dick, G.~C. Gardner, M.~J. Manfra, and D.~J. Reilly.
\newblock ``{Cryogenic Interface for Controlling Many Qubits}''~(2019).
\newblock  \href{http://arxiv.org/abs/1912.01299}{arXiv:1912.01299}.

\bibitem{Li2018a}
Ruoyu Li, Luca Petit, David~P. Franke, Juan~Pablo Dehollain, Jonas Helsen, Mark Steudtner, Nicole~K. Thomas, Zachary~R. Yoscovits, Kanwal~J. Singh, Stephanie Wehner, Lieven M.~K. Vandersypen, James~S. Clarke, and Menno Veldhorst.
\newblock ``{A crossbar network for silicon quantum dot qubits}''.
\newblock \href{https://doi.org/10.1126/sciadv.aar3960}{Science Advances {\bf 4}, 33}~(2018).

\bibitem{Boter2022}
Jelmer~M. Boter, Juan~P. Dehollain, Jeroen~P.G. van Dijk, Yuanxing Xu, Toivo Hensgens, Richard Versluis, Henricus~W.L. Naus, James~S. Clarke, Menno Veldhorst, Fabio Sebastiano, and Lieven~M.K. Vandersypen.
\newblock ``{Spiderweb Array: A Sparse Spin-Qubit Array}''.
\newblock \href{https://doi.org/10.1103/physrevapplied.18.024053}{Physical Review Applied {\bf 18}, 024053}~(2022).

\bibitem{Vandersypen2017}
L.~M.~K. Vandersypen, H.~Bluhm, J.~S. Clarke, A.~S. Dzurak, R.~Ishihara, A.~Morello, D.~J. Reilly, L.~R. Schreiber, and M.~Veldhorst.
\newblock ``{Interfacing spin qubits in quantum dots and donors{\textemdash}hot, dense, and coherent}''.
\newblock \href{https://doi.org/10.1038/s41534-017-0038-y}{npj Quantum Information {\bf 3}, 34}~(2017).

\bibitem{HarveyCollard2022}
Patrick Harvey-Collard, Jurgen Dijkema, Guoji Zheng, Amir Sammak, Giordano Scappucci, and Lieven~M.{\hspace{0.167em}}K. Vandersypen.
\newblock ``{Coherent Spin-Spin Coupling Mediated by Virtual Microwave Photons}''.
\newblock \href{https://doi.org/10.1103/physrevx.12.021026}{Physical Review X {\bf 12}, 021026}~(2022).

\bibitem{Shilton1996}
J~M Shilton, V~I Talyanskii, M~Pepper, D~A Ritchie, J~E~F Frost, C~J~B Ford, C~G Smith, and G~A~C Jones.
\newblock ``{High-frequency single-electron transport in a quasi-one-dimensional {GaAs} channel induced by surface acoustic waves}''.
\newblock \href{https://doi.org/10.1088/0953-8984/8/38/001}{Journal of Physics: Condensed Matter {\bf 8}, L531--L539}~(1996).

\bibitem{Mills2019}
A.~R. Mills, D.~M. Zajac, M.~J. Gullans, F.~J. Schupp, T.~M. Hazard, and J.~R. Petta.
\newblock ``{Shuttling a single charge across a one-dimensional array of silicon quantum dots}''.
\newblock \href{https://doi.org/10.1038/s41467-019-08970-z}{Nature Communications {\bf 10}, 1063}~(2019).

\bibitem{RiggelenDoelman2024}
Floor van Riggelen-Doelman, Chien-An Wang, Sander~L. de~Snoo, William I.~L. Lawrie, Nico~W. Hendrickx, Maximilian Rimbach-Russ, Amir Sammak, Giordano Scappucci, Corentin Déprez, and Menno Veldhorst.
\newblock ``Coherent spin qubit shuttling through germanium quantum dots''.
\newblock \href{https://doi.org/10.1038/s41467-024-49358-y}{Nature Communications {\bf 15}}~(2024).

\bibitem{Zwerver2023}
A.M.J. Zwerver, S.V. Amitonov, S.L. de~Snoo, M.T. Mądzik, M.~Rimbach-Russ, A.~Sammak, G.~Scappucci, and L.M.K. Vandersypen.
\newblock ``Shuttling an electron spin through a silicon quantum dot array''.
\newblock \href{https://doi.org/10.1103/prxquantum.4.030303}{PRX Quantum {\bf 4}, 030303}~(2023).

\bibitem{Seidler2022}
Inga Seidler, Tom Struck, Ran Xue, Niels Focke, Stefan Trellenkamp, Hendrik Bluhm, and Lars~R. Schreiber.
\newblock ``{Conveyor-mode single-electron shuttling in Si/{SiGe} for a scalable quantum computing architecture}''.
\newblock \href{https://doi.org/10.1038/s41534-022-00615-2}{npj Quantum Information {\bf 8}, 100}~(2022).

\bibitem{Struck2024}
Tom Struck, Mats Volmer, Lino Visser, Tobias Offermann, Ran Xue, Jhih-Sian Tu, Stefan Trellenkamp, Łukasz Cywiński, Hendrik Bluhm, and Lars~R. Schreiber.
\newblock ``Spin-epr-pair separation by conveyor-mode single electron shuttling in si/sige''.
\newblock \href{https://doi.org/10.1038/s41467-024-45583-7}{Nature Communications {\bf 15}}~(2024).

\bibitem{Langrock2023}
{Veit Langrock and Jan A. Krzywda and Niels Focke and Inga Seidler and Lars R. Schreiber and {\L}ukasz Cywi{\'{n}}ski}.
\newblock ``{Blueprint of a Scalable Spin Qubit Shuttle Device for Coherent Mid-Range Qubit Transfer in Disordered ${\text{Si/SiGe/SiO}}_{2}$}''.
\newblock \href{https://doi.org/10.1103/prxquantum.4.020305}{{PRX} Quantum {\bf 4}, 020305}~(2023).

\bibitem{Xue2024}
Ran Xue, Max Beer, Inga Seidler, Simon Humpohl, Jhih-Sian Tu, Stefan Trellenkamp, Tom Struck, Hendrik Bluhm, and Lars~R. Schreiber.
\newblock ``Si/sige qubus for single electron information-processing devices with memory and micron-scale connectivity function''.
\newblock \href{https://doi.org/10.1038/s41467-024-46519-x}{Nature Communications {\bf 15}}~(2024).

\bibitem{Buonacorsi2019}
Brandon Buonacorsi, Zhenyu Cai, Eduardo~B Ramirez, Kyle~S Willick, Sean~M Walker, Jiahao Li, Benjamin~D Shaw, Xiaosi Xu, Simon~C Benjamin, and Jonathan Baugh.
\newblock ``{Network architecture for a topological quantum computer in silicon}''.
\newblock \href{https://doi.org/10.1088/2058-9565/aaf3c4}{Quantum Science and Technology {\bf 4}, 025003}~(2019).

\bibitem{Jnane2022}
Hamza Jnane, Brennan Undseth, Zhenyu Cai, Simon~C. Benjamin, and B{\'{a}}lint Koczor.
\newblock ``{Multicore Quantum Computing}''.
\newblock \href{https://doi.org/10.1103/physrevapplied.18.044064}{Physical Review Applied {\bf 18}, 044064}~(2022).

\bibitem{Kuenne2024}
Matthias Künne, Alexander Willmes, Max Oberländer, Christian Gorjaew, Julian~D. Teske, Harsh Bhardwaj, Max Beer, Eugen Kammerloher, René Otten, Inga Seidler, Ran Xue, Lars~R. Schreiber, and Hendrik Bluhm.
\newblock ``The spinbus architecture for scaling spin qubits with electron shuttling''.
\newblock \href{https://doi.org/10.1038/s41467-024-49182-4}{Nature Communications {\bf 15}}~(2024).

\bibitem{flyingQ-photon}
Peter Lodahl.
\newblock ``Quantum-dot based photonic quantum networks''.
\newblock \href{https://doi.org/10.1088/2058-9565/aa91bb}{Quantum Science and Technology {\bf 3}, 013001}~(2017).

\bibitem{flyingQ-electron2}
Christopher Bäuerle, D~Christian Glattli, Tristan Meunier, Fabien Portier, Patrice Roche, Preden Roulleau, Shintaro Takada, and Xavier Waintal.
\newblock ``Coherent control of single electrons: a review of current progress''.
\newblock \href{https://doi.org/10.1088/1361-6633/aaa98a}{Reports on Progress in Physics {\bf 81}, 056503}~(2018).

\bibitem{flyingQ-electron1}
Hermann Edlbauer, Junliang Wang, Thierry Crozes, Pierre Perrier, Seddik Ouacel, Cl{\'e}ment Geffroy, Giorgos Georgiou, Eleni Chatzikyriakou, Antonio Lacerda-Santos, Xavier Waintal, D.~Christian Glattli, Preden Roulleau, Jayshankar Nath, Masaya Kataoka, Janine Splettstoesser, Matteo Acciai, Maria~Cecilia da~Silva~Figueira, Kemal {\"O}ztas, Alex Trellakis, Thomas Grange, Oleg~M. Yevtushenko, Stefan Birner, and Christopher B{\"a}uerle.
\newblock ``Semiconductor-based electron flying qubits: review on recent progress accelerated by numerical modelling''.
\newblock \href{https://doi.org/10.1140/epjqt/s40507-022-00139-w}{EPJ Quantum Technology {\bf 9}, 21}~(2022).

\bibitem{flyingQ-QC}
David~P. DiVincenzo.
\newblock ``The physical implementation of quantum computation''.
\newblock \href{https://doi.org/10.1002/3527603182.ch1}{Fortschritte der Physik {\bf 48}, 771--783}~(2000).

\bibitem{flyingQ-Qinternet}
Stephanie Wehner, David Elkouss, and Ronald Hanson.
\newblock ``Quantum internet: A vision for the road ahead''.
\newblock \href{https://doi.org/10.1126/science.aam9288}{Science {\bf 362}, eaam9288}~(2018).

\bibitem{Greentree2004}
Andrew~D. Greentree, Jared~H. Cole, A.~R. Hamilton, and Lloyd C.~L. Hollenberg.
\newblock ``{Coherent electronic transfer in quantum dot systems using adiabatic passage}''.
\newblock \href{https://doi.org/10.1103/physrevb.70.235317}{Physical Review B {\bf 70}, 235317}~(2004).

\bibitem{Michaelis2006}
B~Michaelis, C~Emary, and C.~W.~J Beenakker.
\newblock ``All-electronic coherent population trapping in quantum dots''.
\newblock \href{https://doi.org/10.1209/epl/i2005-10458-6}{Europhysics Letters (EPL) {\bf 73}, 677--683}~(2006).

\bibitem{Sanchez2013}
Rafael Sánchez and Gloria Platero.
\newblock ``Dark bell states in tunnel-coupled spin qubits''.
\newblock \href{https://doi.org/10.1103/physrevb.87.081305}{Physical Review B {\bf 87}, 081305(R)}~(2013).

\bibitem{Ban2018}
Yue Ban, Xi~Chen, and Gloria Platero.
\newblock ``{Fast long-range charge transfer in quantum dot arrays}''.
\newblock \href{https://doi.org/10.1088/1361-6528/aae0ce}{Nanotechnology {\bf 29}, 505201}~(2018).

\bibitem{Ban2019}
Yue Ban, Xi~Chen, Sigmund Kohler, and Gloria Platero.
\newblock ``{Spin Entangled State Transfer in Quantum Dot Arrays: Coherent Adiabatic and Speed-Up Protocols}''.
\newblock \href{https://doi.org/10.1002/qute.201900048}{Advanced Quantum Technologies {\bf 2}, 1900048}~(2019).

\bibitem{Gullans2020}
M.~J. Gullans and J.~R. Petta.
\newblock ``{Coherent transport of spin by adiabatic passage in quantum dot arrays}''.
\newblock \href{https://doi.org/10.1103/physrevb.102.155404}{Physical Review B {\bf 102}, 155404}~(2020).

\bibitem{Bogan2017}
A.~Bogan, S.~A. Studenikin, M.~Korkusinski, G.~C. Aers, L.~Gaudreau, P.~Zawadzki, A.~S. Sachrajda, L.~A. Tracy, J.~L. Reno, and T.~W. Hargett.
\newblock ``Consequences of spin-orbit coupling at the single hole level: Spin-flip tunneling and the anisotropic $g$ factor''.
\newblock \href{https://doi.org/10.1103/PhysRevLett.118.167701}{Phys. Rev. Lett. {\bf 118}, 167701}~(2017).

\bibitem{Bogan2018}
Alex Bogan, Sergei Studenikin, Marek Korkusinski, Louis Gaudreau, Piotr Zawadzki, Andy~S. Sachrajda, Lisa Tracy, John Reno, and Terry Hargett.
\newblock ``{Landau-Zener-Stückelberg-Majorana Interferometry of a Single Hole}''.
\newblock \href{https://doi.org/10.1103/physrevlett.120.207701}{Physical Review Letters {\bf 120}, 207701}~(2018).

\bibitem{Studenikin2019}
Sergei Studenikin, Marek Korkusinski, Motoi Takahashi, Jordan Ducatel, Aviv Padawer-Blatt, Alex Bogan, D.~Guy Austing, Louis Gaudreau, Piotr Zawadzki, Andrew Sachrajda, Yoshiro Hirayama, Lisa Tracy, John Reno, and Terry Hargett.
\newblock ``Electrically tunable effective g-factor of a single hole in a lateral gaas/algaas quantum dot''.
\newblock \href{https://doi.org/10.1038/s42005-019-0262-1}{Communications Physics {\bf 2}, 159}~(2019).

\bibitem{Bogan2019}
Alex Bogan, Sergei Studenikin, Marek Korkusinski, Louis Gaudreau, Piotr Zawadzki, Andy Sachrajda, Lisa Tracy, John Reno, and Terry Hargett.
\newblock ``Single hole spin relaxation probed by fast single-shot latched charge sensing''.
\newblock \href{https://doi.org/10.1038/s42005-019-0113-0}{Communications Physics {\bf 2}, 17}~(2019).

\bibitem{Studenikin2021}
Sergei Studenikin, Marek Korkusinski, Alex Bogan, Louis Gaudreau, D~Guy Austing, Andrew~S Sachrajda, Lisa Tracy, John Reno, and Terry Hargett.
\newblock ``Single-hole physics in gaas/algaas double quantum dot system with strong spin–orbit interaction''.
\newblock \href{https://doi.org/10.1088/1361-6641/abe42d}{Semiconductor Science and Technology {\bf 36}, 053001}~(2021).

\bibitem{Ducatel2021}
J.~Ducatel, A.~Padawer-Blatt, A.~Bogan, M.~Korkusinski, P.~Zawadzki, A.~Sachrajda, S.~Studenikin, L.~Tracy, J.~Reno, and T.~Hargett.
\newblock ``Single-hole couplings in gaas/algaas double dots probed with transport and edsr spectroscopy''.
\newblock \href{https://doi.org/10.1063/5.0044933}{Applied Physics Letters {\bf 118}, 214002}~(2021).

\bibitem{Bogan2021}
Alex Bogan, Sergei Studenikin, Marek Korkusinski, Louis Gaudreau, Jason Phoenix, Piotr Zawadzki, Andy Sachrajda, Lisa Tracy, John Reno, and Terry Hargett.
\newblock ``Spin-orbit enabled quantum transport channels in a two-hole double quantum dot''.
\newblock \href{https://doi.org/10.1103/physrevb.103.235310}{Physical Review B {\bf 103}, 235310}~(2021).

\bibitem{PadawerBlatt2022}
A.~Padawer-Blatt, J.~Ducatel, M.~Korkusinski, A.~Bogan, L.~Gaudreau, P.~Zawadzki, D.~G. Austing, A.~S. Sachrajda, S.~Studenikin, L.~Tracy, J.~Reno, and T.~Hargett.
\newblock ``Characterization of dot-specific and tunable effective $g$ factors in a gaas/algaas double quantum dot single-hole device''.
\newblock \href{https://doi.org/10.1103/physrevb.105.195305}{Physical Review B {\bf 105}, 195305}~(2022).

\bibitem{Marton2023}
Victor Marton, Andrew Sachrajda, Marek Korkusinski, Alex Bogan, and Sergei Studenikin.
\newblock ``Coherence characteristics of a gaas single heavy-hole spin qubit using a modified single-shot latching readout technique''.
\newblock \href{https://doi.org/10.3390/nano13050950}{Nanomaterials {\bf 13}, 950}~(2023).

\bibitem{Ginzel2020}
Florian Ginzel, Adam~R. Mills, Jason~R. Petta, and Guido Burkard.
\newblock ``Spin shuttling in a silicon double quantum dot''.
\newblock \href{https://doi.org/10.1103/physrevb.102.195418}{Physical Review B {\bf 102}, 195418}~(2020).

\bibitem{Qi2023}
Jiaan Qi, Zhi-Hai Liu, and Hongqi Xu.
\newblock ``Spin–orbit interaction enabled high-fidelity two-qubit gates''.
\newblock \href{https://doi.org/10.1088/1367-2630/ad19ab}{New Journal of Physics {\bf 26}, 013012}~(2024).

\bibitem{Liu2024}
Xiao-Fei Liu, Yuta Matsumoto, Takafumi Fujita, Arne Ludwig, Andreas~D. Wieck, and Akira Oiwa.
\newblock ``Accelerated adiabatic passage of a single electron spin qubit in quantum dots''.
\newblock \href{https://doi.org/10.1103/physrevlett.132.027002}{Physical Review Letters {\bf 132}, 027002}~(2024).

\bibitem{GueryOdelin2019}
D.~Gu{\'{e}}ry-Odelin, A.~Ruschhaupt, A.~Kiely, E.~Torrontegui, S.~Mart{\'{\i}}nez-Garaot, and J.{\hspace{0.167em}}G. Muga.
\newblock ``{Shortcuts to adiabaticity: Concepts, methods, and applications}''.
\newblock \href{https://doi.org/10.1103/revmodphys.91.045001}{Reviews of Modern Physics {\bf 91}, 045001}~(2019).

\bibitem{Zajac2016}
D.{\hspace{0.167em}}M. Zajac, T.{\hspace{0.167em}}M. Hazard, X.~Mi, E.~Nielsen, and J.{\hspace{0.167em}}R. Petta.
\newblock ``{Scalable Gate Architecture for a One-Dimensional Array of Semiconductor Spin Qubits}''.
\newblock \href{https://doi.org/10.1103/physrevapplied.6.054013}{Physical Review Applied {\bf 6}, 054013}~(2016).

\bibitem{Otsuka2016}
Tomohiro Otsuka, Takashi Nakajima, Matthieu~R. Delbecq, Shinichi Amaha, Jun Yoneda, Kenta Takeda, Giles Allison, Takumi Ito, Retsu Sugawara, Akito Noiri, Arne Ludwig, Andreas~D. Wieck, and Seigo Tarucha.
\newblock ``{Single-electron Spin Resonance in a Quadruple Quantum Dot}''.
\newblock \href{https://doi.org/10.1038/srep31820}{Scientific Reports {\bf 6}, 31820}~(2016).

\bibitem{Baart2016}
T.~A. Baart, M.~Shafiei, T.~Fujita, C.~Reichl, W.~Wegscheider, and L.~M.~K. Vandersypen.
\newblock ``{Single-spin {CCD}}''.
\newblock \href{https://doi.org/10.1038/nnano.2015.291}{Nature Nanotechnology {\bf 11}, 330--334}~(2016).

\bibitem{Fujita2017}
Takafumi Fujita, Timothy~Alexander Baart, Christian Reichl, Werner Wegscheider, and Lieven Mark~Koenraad Vandersypen.
\newblock ``{Coherent shuttle of electron-spin states}''.
\newblock \href{https://doi.org/10.1038/s41534-017-0024-4}{npj Quantum Information {\bf 3}, 22}~(2017).

\bibitem{Kandel2019}
Yadav~P. Kandel, Haifeng Qiao, Saeed Fallahi, Geoffrey~C. Gardner, Michael~J. Manfra, and John~M. Nichol.
\newblock ``{Coherent spin-state transfer via Heisenberg exchange}''.
\newblock \href{https://doi.org/10.1038/s41586-019-1566-8}{Nature {\bf 573}, 553--557}~(2019).

\bibitem{Qiao2020}
Haifeng Qiao, Yadav~P. Kandel, Sreenath~K. Manikandan, Andrew~N. Jordan, Saeed Fallahi, Geoffrey~C. Gardner, Michael~J. Manfra, and John~M. Nichol.
\newblock ``Conditional teleportation of quantum-dot spin states''.
\newblock \href{https://doi.org/10.1038/s41467-020-16745-0}{Nature Communications {\bf 11}, 3022}~(2020).

\bibitem{Lawrie2020}
W.~I.~L. Lawrie, H.~G.~J. Eenink, N.~W. Hendrickx, J.~M. Boter, L.~Petit, S.~V. Amitonov, M.~Lodari, B.~Paquelet Wuetz, C.~Volk, S.~G.~J. Philips, G.~Droulers, N.~Kalhor, F.~van Riggelen, D.~Brousse, A.~Sammak, L.~M.~K. Vandersypen, G.~Scappucci, and M.~Veldhorst.
\newblock ``{Quantum dot arrays in silicon and germanium}''.
\newblock \href{https://doi.org/10.1063/5.0002013}{Applied Physics Letters {\bf 116}, 080501}~(2020).

\bibitem{Yu2023}
C{\'{e}}cile~X. Yu, Simon Zihlmann, Jos{\'{e}}~C. Abadillo-Uriel, Vincent~P. Michal, Nils Rambal, Heimanu Niebojewski, Thomas Bedecarrats, Maud Vinet, {\'{E}}tienne Dumur, Michele Filippone, Benoit Bertrand, Silvano~De Franceschi, Yann-Michel Niquet, and Romain Maurand.
\newblock ``{Strong coupling between a photon and a hole spin in silicon}''.
\newblock \href{https://doi.org/10.1038/s41565-023-01332-3}{Nature Nanotechnology {\bf 18}, 741--746}~(2023).

\bibitem{Fang2023}
Yinan Fang, Pericles Philippopoulos, Dimitrie Culcer, W~A Coish, and Stefano Chesi.
\newblock ``Recent advances in hole-spin qubits''.
\newblock \href{https://doi.org/10.1088/2633-4356/acb87e}{Materials for Quantum Technology {\bf 3}, 012003}~(2023).

\bibitem{FernandezFernandez2022}
D.~Fern{\'{a}}ndez-Fern{\'{a}}ndez, Yue Ban, and G.~Platero.
\newblock ``{Quantum Control of Hole Spin Qubits in Double Quantum Dots}''.
\newblock \href{https://doi.org/10.1103/physrevapplied.18.054090}{Physical Review Applied {\bf 18}, 054090}~(2022).

\bibitem{FernandezFernandez2023}
David Fern{\'{a}}ndez-Fern{\'{a}}ndez, Jordi Pic{\'{o}}-Cort{\'{e}}s, Sergio~Vela Li{\~{n}}{\'{a}}n, and Gloria Platero.
\newblock ``{Photo-assisted spin transport in double quantum dots with spin{\textendash}orbit interaction}''.
\newblock \href{https://doi.org/10.1088/2515-7639/acd1b7}{Journal of Physics: Materials {\bf 6}, 034004}~(2023).

\bibitem{Burkard2023}
Guido Burkard, Thaddeus~D. Ladd, Andrew Pan, John~M. Nichol, and Jason~R. Petta.
\newblock ``Semiconductor spin qubits''.
\newblock \href{https://doi.org/10.1103/revmodphys.95.025003}{Reviews of Modern Physics {\bf 95}, 025003}~(2023).

\bibitem{Mutter2020}
Philipp~M. Mutter and Guido Burkard.
\newblock ``{Cavity control over heavy-hole spin qubits in inversion-symmetric crystals}''.
\newblock \href{https://doi.org/10.1103/physrevb.102.205412}{Physical Review B {\bf 102}, 205412}~(2020).

\bibitem{Bosco2021}
Stefano Bosco, M{\'{o}}nica Benito, Christoph Adelsberger, and Daniel Loss.
\newblock ``{Squeezed hole spin qubits in Ge quantum dots with ultrafast gates at low power}''.
\newblock \href{https://doi.org/10.1103/physrevb.104.115425}{Physical Review B {\bf 104}, 115425}~(2021).

\bibitem{Mutter2021b}
Philipp~M. Mutter and Guido Burkard.
\newblock ``{All-electrical control of hole singlet-triplet spin qubits at low-leakage points}''.
\newblock \href{https://doi.org/10.1103/physrevb.104.195421}{Physical Review B {\bf 104}, 195421}~(2021).

\bibitem{Mutter2021a}
Philipp~M. Mutter and Guido Burkard.
\newblock ``{Natural heavy-hole flopping mode qubit in germanium}''.
\newblock \href{https://doi.org/10.1103/physrevresearch.3.013194}{Physical Review Research {\bf 3}, 013194}~(2021).

\bibitem{Adelsberger2022}
Christoph Adelsberger, M{\'{o}}nica Benito, Stefano Bosco, Jelena Klinovaja, and Daniel Loss.
\newblock ``{Hole-spin qubits in Ge nanowire quantum dots: Interplay of orbital magnetic field, strain, and growth direction}''.
\newblock \href{https://doi.org/10.1103/physrevb.105.075308}{Physical Review B {\bf 105}, 075308}~(2022).

\bibitem{Jirovec2022}
Daniel Jirovec, Philipp~M. Mutter, Andrea Hofmann, Alessandro Crippa, Marek Rychetsky, David~L. Craig, Josip Kukucka, Frederico Martins, Andrea Ballabio, Natalia Ares, Daniel Chrastina, Giovanni Isella, Guido Burkard, and Georgios Katsaros.
\newblock ``{Dynamics of Hole Singlet-Triplet Qubits with Large $g$-Factor Differences}''.
\newblock \href{https://doi.org/10.1103/physrevlett.128.126803}{Physical Review Letters {\bf 128}, 126803}~(2022).

\bibitem{Nitta1997}
Junsaku Nitta, Tatsushi Akazaki, Hideaki Takayanagi, and Takatomo Enoki.
\newblock ``Gate control of spin-orbit interaction in an inverted i${\mathrm{n}}_{0.53}$g${\mathrm{a}}_{0.47}$as/i${\mathrm{n}}_{0.52}$a${\mathrm{l}}_{0.48}$as heterostructure''.
\newblock \href{https://doi.org/10.1103/PhysRevLett.78.1335}{Phys. Rev. Lett. {\bf 78}, 1335--1338}~(1997).

\bibitem{Mutter2021}
Philipp~M. Mutter and Guido Burkard.
\newblock ``{Pauli spin blockade with site-dependent g tensors and spin-polarized leads}''.
\newblock \href{https://doi.org/10.1103/physrevb.103.245412}{Physical Review B {\bf 103}, 245412}~(2021).

\bibitem{Chen2010}
Xi~Chen, A.~Ruschhaupt, S.~Schmidt, A.~del Campo, D.~Gu{\'{e}}ry-Odelin, and J.~G. Muga.
\newblock ``{Fast Optimal Frictionless Atom Cooling in Harmonic Traps: Shortcut to Adiabaticity}''.
\newblock \href{https://doi.org/10.1103/physrevlett.104.063002}{Physical Review Letters {\bf 104}, 063002}~(2010).

\bibitem{Malinovsky1997}
Vladimir~S. Malinovsky and David~J. Tannor.
\newblock ``Simple and robust extension of the stimulated raman adiabatic passage technique to $n$-level systems''.
\newblock \href{https://doi.org/10.1103/physreva.56.4929}{Physical Review A {\bf 56}, 4929--4937}~(1997).

\bibitem{Vitanov2001a}
N.V. Vitanov, M.~Fleischhauer, B.W. Shore, and K.~Bergmann.
\newblock ``Coherent manipulation of atoms molecules by sequential laser pulses''.
\newblock In Benjamin Bederson and Herbert Walther, editors, Advances In Atomic, Molecular, and Optical Physics.
\newblock \href{https://doi.org/10.1016/s1049-250x(01)80063-x}{Volume~46, pages 55--190}.
\newblock Academic Press~(2001).

\bibitem{Manzano2020}
Daniel Manzano.
\newblock ``A short introduction to the lindblad master equation''.
\newblock \href{https://doi.org/10.1063/1.5115323}{AIP Advances {\bf 10}, 025106}~(2020).

\bibitem{Malshukov2003}
A.~G. Mal'shukov, C.~S. Tang, C.~S. Chu, and K.~A. Chao.
\newblock ``{Spin-current generation and detection in the presence of an ac gate}''.
\newblock \href{https://doi.org/10.1103/physrevb.68.233307}{Physical Review B {\bf 68}, 233307}~(2003).

\bibitem{Faniel2011}
S.~Faniel, T.~Matsuura, S.~Mineshige, Y.~Sekine, and T.~Koga.
\newblock ``{Determination of spin-orbit coefficients in semiconductor quantum wells}''.
\newblock \href{https://doi.org/10.1103/physrevb.83.115309}{Physical Review B {\bf 83}, 115309}~(2011).

\bibitem{Li2018}
Yi-Chao Li, Xi~Chen, J~G Muga, and E~Ya Sherman.
\newblock ``{Qubit gates with simultaneous transport in double quantum dots}''.
\newblock \href{https://doi.org/10.1088/1367-2630/aaedd9}{New Journal of Physics {\bf 20}, 113029}~(2018).

\bibitem{Barenco1995}
Adriano Barenco, Charles~H. Bennett, Richard Cleve, David~P. DiVincenzo, Norman Margolus, Peter Shor, Tycho Sleator, John~A. Smolin, and Harald Weinfurter.
\newblock ``Elementary gates for quantum computation''.
\newblock \href{https://doi.org/10.1103/physreva.52.3457}{Physical Review A {\bf 52}, 3457--3467}~(1995).

\bibitem{Shim2013}
Yun-Pil Shim, Jianjia Fei, Sangchul Oh, Xuedong Hu, and Mark Friesen.
\newblock ``Single-qubit gates in two steps with rotation axes in a single plane''~(2013).
\newblock  \href{http://arxiv.org/abs/1303.0297}{arXiv:1303.0297}.

\bibitem{Stano2022}
Peter Stano and Daniel Loss.
\newblock ``Review of performance metrics of spin qubits in gated semiconducting nanostructures''.
\newblock \href{https://doi.org/10.1038/s42254-022-00484-w}{Nature Reviews Physics {\bf 4}, 672--688}~(2022).

\bibitem{Huang2019}
W.~Huang, C.~H. Yang, K.~W. Chan, T.~Tanttu, B.~Hensen, R.~C.~C. Leon, M.~A. Fogarty, J.~C.~C. Hwang, F.~E. Hudson, K.~M. Itoh, A.~Morello, A.~Laucht, and A.~S. Dzurak.
\newblock ``{Fidelity benchmarks for two-qubit gates in silicon}''.
\newblock \href{https://doi.org/10.1038/s41586-019-1197-0}{Nature {\bf 569}, 532--536}~(2019).

\bibitem{Hendrickx2021}
Nico~W. Hendrickx, William I.~L. Lawrie, Maximilian Russ, Floor van Riggelen, Sander~L. de~Snoo, Raymond~N. Schouten, Amir Sammak, Giordano Scappucci, and Menno Veldhorst.
\newblock ``{A four-qubit germanium quantum processor}''.
\newblock \href{https://doi.org/10.1038/s41586-021-03332-6}{Nature {\bf 591}, 580--585}~(2021).

\bibitem{Assali2011}
Lucy V.~C. Assali, Helena~M. Petrilli, Rodrigo~B. Capaz, Belita Koiller, Xuedong Hu, and S.~Das Sarma.
\newblock ``Hyperfine interactions in silicon quantum dots''.
\newblock \href{https://doi.org/10.1103/physrevb.83.165301}{Physical Review B {\bf 83}, 165301}~(2011).

\bibitem{Carr1954}
H.~Y. Carr and E.~M. Purcell.
\newblock ``Effects of diffusion on free precession in nuclear magnetic resonance experiments''.
\newblock \href{https://doi.org/10.1103/physrev.94.630}{Physical Review {\bf 94}, 630--638}~(1954).

\bibitem{Meiboom1958}
S.~Meiboom and D.~Gill.
\newblock ``Modified spin-echo method for measuring nuclear relaxation times''.
\newblock \href{https://doi.org/10.1063/1.1716296}{Review of Scientific Instruments {\bf 29}, 688--691}~(1958).

\bibitem{Vitanov2017}
Nikolay~V. Vitanov, Andon~A. Rangelov, Bruce~W. Shore, and Klaas Bergmann.
\newblock ``{Stimulated Raman adiabatic passage in physics, chemistry, and beyond}''.
\newblock \href{https://doi.org/10.1103/revmodphys.89.015006}{Reviews of Modern Physics {\bf 89}, 015006}~(2017).

\bibitem{Rashba1988}
E.I. Rashba and E.Ya. Sherman.
\newblock ``Spin-orbital band splitting in symmetric quantum wells''.
\newblock \href{https://doi.org/10.1016/0375-9601(88)90140-5}{Physics Letters A {\bf 129}, 175--179}~(1988).

\bibitem{Luo2010}
Jun-Wei Luo, Athanasios~N. Chantis, Mark van Schilfgaarde, Gabriel Bester, and Alex Zunger.
\newblock ``Discovery of a novel linear-in-$k$ spin splitting for holes in the 2d $\mathrm{GaAs}/\mathrm{AlAs}$ system''.
\newblock \href{https://doi.org/10.1103/physrevlett.104.066405}{Physical Review Letters {\bf 104}, 066405}~(2010).

\bibitem{Szumniak2012}
P.~Szumniak, S.~Bednarek, B.~Partoens, and F.~M. Peeters.
\newblock ``{Spin-Orbit-Mediated Manipulation of Heavy-Hole Spin Qubits in Gated Semiconductor Nanodevices}''.
\newblock \href{https://doi.org/10.1103/physrevlett.109.107201}{Physical Review Letters {\bf 109}, 107201}~(2012).

\bibitem{Liu2022}
Yang Liu, Jia-Xin Xiong, Zhi Wang, Wen-Long Ma, Shan Guan, Jun-Wei Luo, and Shu-Shen Li.
\newblock ``{Emergent linear Rashba spin-orbit coupling offers fast manipulation of hole-spin qubits in germanium}''.
\newblock \href{https://doi.org/10.1103/physrevb.105.075313}{Physical Review B {\bf 105}, 075313}~(2022).

\bibitem{Altarelli1988}
M.~Altarelli and G.~Platero.
\newblock ``Magnetic hole levels in quantum wells in a parallel field''.
\newblock \href{https://doi.org/10.1016/0039-6028(88)90738-8}{Surface Science {\bf 196}, 540--544}~(1988).

\bibitem{Platero1989}
G.~Platero and M.~Altarelli.
\newblock ``Valence-band levels and optical transitions in quantum wells in a parallel magnetic field''.
\newblock \href{https://doi.org/10.1103/physrevb.39.3758}{Physical Review B {\bf 39}, 3758--3763}~(1989).

\bibitem{Luttinger1955}
J.~M. Luttinger and W.~Kohn.
\newblock ``Motion of electrons and holes in perturbed periodic fields''.
\newblock \href{https://doi.org/10.1103/physrev.97.869}{Physical Review {\bf 97}, 869--883}~(1955).

\bibitem{Winkler2003}
Roland Winkler.
\newblock ``Spin—orbit coupling effects in two-dimensional electron and hole systems''.
\newblock \href{https://doi.org/10.1007/b13586}{Springer Berlin Heidelberg}. ~(2003).

\bibitem{Gullans2019}
M.~J. Gullans and J.~R. Petta.
\newblock ``{Protocol for a resonantly driven three-qubit Toffoli gate with silicon spin qubits}''.
\newblock \href{https://doi.org/10.1103/physrevb.100.085419}{Physical Review B {\bf 100}, 085419}~(2019).

\end{thebibliography}

\end{document}